\documentclass{article}

\usepackage{arxiv}

\usepackage[utf8]{inputenc} 
\usepackage[T1]{fontenc}    
\usepackage{hyperref}       
\usepackage{url}            
\usepackage{booktabs}       
\usepackage{amsfonts}       
\usepackage{nicefrac}       
\usepackage{microtype}      
\usepackage{lipsum}		
\usepackage{graphicx}
\usepackage[square, comma]{natbib}
\usepackage{doi}
\usepackage{graphicx} 
\usepackage{geometry}
\usepackage{enumitem}
\usepackage{appendix}
\usepackage{url}
\usepackage{xcolor}
\usepackage{amsmath}
\usepackage{subfig}
\usepackage{makecell}
\usepackage{multirow}
\usepackage{fancybox}
\usepackage[ruled,vlined]{algorithm2e}
\SetKwProg{Proc}{Procedure}{}{end}
\usepackage{natbib}
\usepackage{longtable}
\usepackage{rotating}

\definecolor{mygrey}{RGB}{230,230,230}
\newsavebox\CBox
\def\textBF#1{\textbf{\texttt{#1}}}

\title{User Intent Recognition and Satisfaction with Large Language Models: A User Study with ChatGPT}

\date{January 2024}

\author{\hspace{1mm}Anna~Bodonhelyi\\
	Chair for Human-Centered Technologies for Learning\\
	Technical University of Munich\\
	Munich, Bavaria, Germany \\
	\texttt{anna.bodonhelyi@tum.de} \\
	\And
	\hspace{1mm}Efe~Bozkir \\
	Chair for Human-Centered Technologies for Learning\\
	Technical University of Munich\\
	Munich, Bavaria, Germany \\
	\texttt{efe.bozkir@tum.de} \\
        \And
	\hspace{1mm}Shuo~Yang \\
	Chair for Responsible Data Science\\
	Technical University of Munich\\
	Munich, Bavaria, Germany \\
	\texttt{shuo.yang@tum.de} \\
        \And
	\hspace{1mm}Enkelejda~Kasneci \\
	Chair for Human-Centered Technologies for Learning\\
	Technical University of Munich\\
	Munich, Bavaria, Germany \\
	\texttt{enkelejda.kasneci@tum.de} \\
        \And
	\hspace{1mm}Gjergji~Kasneci \\
	Chair for Responsible Data Science\\
	Technical University of Munich\\
	Munich, Bavaria, Germany \\
	\texttt{gjergji.kasneci@tum.de} \\
}

\renewcommand{\shorttitle}{\textit{arXiv} Template}

\hypersetup{
pdftitle={User Intent Recognition and Satisfaction with Large Language Models: A User Study with ChatGPT},
pdfsubject={Computer Science, Human-Computer Interaction},
pdfauthor={Anna Bodonhelyi, Efe Bozkir, Shuo Yang, Enkelejda Kasneci, Gjergji Kasneci},
pdfkeywords={large language models, ChatGPT, user study, intent recognition, prompt reformulation, intent taxonomy, human-AI interaction},
colorlinks=true,
linkcolor=black,
    filecolor=black,      
    urlcolor=blue,
    citecolor=black,
}

\begin{document}
\maketitle

\begin{abstract}
The rapid evolution of LLMs represents an impactful paradigm shift in digital interaction and content engagement. While they encode vast amounts of human-generated knowledge and excel in processing diverse data types, they often face the challenge of accurately responding to specific user intents, leading to user dissatisfaction. Based on a fine-grained intent taxonomy and intent-based prompt reformulations, we analyze the quality of intent recognition and user satisfaction with answers from intent-based prompt reformulations of GPT-3.5 Turbo and GPT-4 Turbo models. Our study highlights the importance of human-AI interaction and underscores the need for interdisciplinary approaches to improve conversational AI systems. We show that GPT-4 outperforms GPT-3.5 in recognizing common intents but is often outperformed by GPT-3.5 in recognizing less frequent intents. Moreover, whenever the user intent is correctly recognized, while users are more satisfied with the intent-based reformulations of GPT-4 compared to GPT-3.5, they tend to be more satisfied with the models' answers to their original prompts compared to the reformulated ones. The collected data from our study has been made publicly available on~\href{https://github.com/ConcealedIDentity/UserIntentStudy}{GitHub} for further research.
\end{abstract}


\section{Introduction}
Generative AI models are rapidly evolving, especially those handling language and vision. Models like OpenAI's GPT-4 Turbo~\citep{OpenAI2023, openai2023gpt4v} and Google's Gemini~\citep{Deepmind2023} are at the forefront of this evolution, impacting how we interact with digital content and services. At the heart of this development are Large Language Models (LLMs). With human-like text processing, generation, and reasoning capabilities, LLMs have broad applications ranging from creative content generation to complex problem-solving~\citep{Brown2020, romera2023mathematical}. Incorporating human feedback and reinforcement learning in the training of LLMs, such as those used in ChatGPT, has further improved the alignment of these models with societal norms and goals~\citep{Ouyang2022, lee2023rlaif, yang2024crowdsourcing}. Similarly, models like Google's PaLM~\citep{Chowdhery2022,Anil2023} and Meta’s LLaMA~\citep{Touvron2023,Touvron2023a}, together with open-source variants like Vicuna~\citep{Zheng2023} and Alpaca~\citep{alpaca2023}, represent parallel advances in the field.

Despite serious challenges in areas like mathematical reasoning~\citep{Imani2023}, student error detection~\citep{bewersdorff2023assessing}, and mitigating hallucinations in outputs~\citep{Ji2023,azamfirei2023large,zhang2023siren,manakul2023selfcheckgpt}, LLMs have already had impact in diverse sectors such as healthcare~\citep{Chintagunta2021,Enarvi2020}, finance~\citep{Dowling2023}, journalism~\citep{Pavlik2023}, creative writing~\citep{Yuan2022, Sessler2023}, and, more recently, to scientific discoveries~\citep{romera2023mathematical}. Efforts towards enhancing explainability are underway to foster trust and reliability in these models~\citep{Wu2023b}. 

Building on the foundations of LLMs, multimodal generative AI models have further expanded the scope to encompass visual, auditory, and other sensory data~\citep{Alayrac2022,Zhu2023,Huang2023a,Ye2023,Li2023,Maaz2023,Driess2023,Chen2023,Su2023,Wu2023,Shen2023}. Hence, the capabilities of LLMs to reason over natural language are essential for many of the recent breakthroughs in generative AI, such as NExT-GPT~\citep{Wu2023a}, OpenAI's GPT-4 Vision, GPT-4 Turbo, Google's Gemini, and Apple's Ferret~\citep{OpenAI2023,Deepmind2023,you2023ferret}. Such advancements, coupled with other deep learning and explainability techniques~\citep{borisov2022deep,rombach2022high}, can potentially revolutionize science~\citep{wong2023discovery} and society. 

Trained on extensive human-generated datasets and various web corpora, LLMs encode extensive knowledge and excel in processing and reasoning tasks over text. Yet, the effectiveness of their responses strongly depends on accurately inferring the user's intent, which is typically conveyed implicitly in the prompt. Accurately discerning and categorizing user intents in prompts poses considerable challenges for LLMs, mainly related to the inherent ambiguity, variability, or the clarity of the used language, cultural and contextual subtleties, and evolving user expectations. A recent study~\citep{kim2023understanding} highlighted a frequent source of user dissatisfaction with ChatGPT: \textbf{its occasional failure to understand user intentions}. The authors further observed that users often lack effective strategies to mitigate this dissatisfaction. Moreover, it was noted that users with a limited understanding of LLMs tend to experience greater dissatisfaction and are less proactive in addressing these challenges. 

In this paper, we explore the hypothesis that enhancing the understanding of a user's specific intent in their prompt can significantly improve the quality of responses generated by Large Language Models (LLMs). This assumption finds substantial backing in the recent advancements in natural language processing methodologies, notably the Chain-of-Thought, Tree-of-Thought, and Graph-of-Thought techniques~\citep{wei2022chain,yao2023tree,besta2023graph}. These approaches emphasize the importance of contextual and hierarchical understanding in processing user queries, which is crucial for effective human-computer interactions. To systematically approach this hypothesis, we first establish a comprehensive taxonomy of potential user intents. This taxonomy is carefully crafted, taking into account the distinct requirements of different intent types and by comprehensively integrating insights from recent and earlier interdisciplinary scholarly work on intent categorization~\citep{dang2022prompt, shah2023using,MediumChatGPTStats, SparktoroChatGPT} and well-established intent categories from information retrieval~\citep{azad2019query, broder2002taxonomy, kellar2007field, jansen2008determining, rose2004understanding, ashkan2009classifying, yi2008deciphering}. By situating our study within the interdisciplinary field of human-computer interaction (HCI), this work examines whether improved intent recognition can enhance user satisfaction and engagement with conversational AI systems.

Our study then proceeds in two critical phases. First, we evaluate recent LLM versions, namely GPT-3.5 and GPT-4, to determine their proficiency in accurately recognizing the intent behind user prompts. We expect more recent LLM versions to outperform older ones on this task. This aspect is crucial, as correctly identifying intent is foundational to providing relevant and accurate responses. The second phase of our study focuses on the ramifications of prompt reformulation. Here, we explore whether accurately recognizing and then reformulating a user's prompt to better mirror their intended query leads to a significant improvement in user satisfaction with the quality of the LLM's responses. 

In brief, our work provides the following contributions:

\begin{enumerate}
    \item The development of a taxonomy of potential intents in conversational contexts.
    \item Assessment of the intent recognition proficiency of GPT-3.5 and GPT-4 based on a user study.
    \item Exploration of the impact of prompt reformulation on response quality based on user satisfaction in a conversational context.
    \item An extensive data set, including an overview of our user study, is published on~\href{https://github.com/ConcealedIDentity/UserIntentStudy}{GitHub} for research purposes.

\end{enumerate}

Section~\ref{sec:related-work} presents existing literature on intent recognition and prompt optimization. We introduce our intent taxonomy in Section~\ref{sec:taxonomy}, laying the groundwork for the user study in Section~\ref{sec:study_description}. The study results are presented and discussed in Section~\ref{sec:results}, before concluding in Section~\ref{sec:conclusion}.

\section{Intent Recognition in User Prompts and Prompt Optimization}
\label{sec:related-work}

This section covers two key aspects of related work: first, intent recognition in user prompts, reviewing methods developed to understand the user's intent in interactions with LLMs; and second, prompt optimization, focusing on the latest techniques in prompt engineering to improve user-LLM interactions. These two research directions are important to emphasize the complex interplay between the user's intent and the model's response. Both aspects are crucial in the field of human-computer interaction, as they directly impact the effectiveness and user satisfaction of conversational AI systems.

\paragraph{Intent recognition in user prompts:} The field of Information Retrieval (IR) has thoroughly examined the concept of intent, leading to various methodologies for understanding and representing it. Search intent classification enables systems to better understand and respond to user objectives. Several of these classifications have been suggested in the literature~\citep{broder2002taxonomy,jansen2007determining}, and taxonomies were developed through iterative processes, primarily involving the manual analysis of search log data. Common categories identified from these logs include: \textit{navigational}, \textit{informational}, \textit{transactional}, \textit{browsing}, and \textit{resource-seeking} intents. However, with the advent of LLMs, the types and dynamics of user interactions are evolving, particularly in terms of \textit{content generation} as opposed to traditional search methods. Consequently, there is ongoing research on new intent classifications tailored to represent intents in these specific interaction contexts. Our research builds upon a recent work~\citep{shah2023using}, where the authors introduced an LLM-based approach with human-in-the-loop to generate intent taxonomies. They introduced five intent categories, namely \emph{information retrieval, problem solving, learning, content creation}, and \emph{leisure}. Their main contribution is the methodology for generating, validating, and using taxonomies for identifying user intents. In contrast, our study focuses on developing a detailed intent classification, including multilevel taxonomies, to enable a more precise understanding and handling of user intents, which is crucial for applications requiring higher granularity. Our work builds upon and significantly advances the foundational efforts of~\citep{shah2023using}, introducing a comprehensive and nuanced taxonomy of user intents tailored to capture the evolving landscape of interactions with large language models (LLMs). To achieve this, we integrate a wide range of user intents drawn from both recent studies and earlier foundational and interdisciplinary literature~\citep{shah2023using, dang2022prompt,azad2019query,MediumChatGPTStats, SparktoroChatGPT, broder2002taxonomy, kellar2007field, jansen2008determining, rose2004understanding, ashkan2009classifying, yi2008deciphering}.

\paragraph{Prompt optimization}
An emerging field in natural language processing is \textit{prompt engineering}, which focuses on designing and optimizing prompts to effectively interact with and guide LLMs. Various techniques have been developed in this domain to enhance the quality of interactions~\citep{liu2023pre}. 
Prompt engineering explores adding text or vectors to inputs and outputs of LLMs to streamline interactions without altering the core parameters of the model, offering thus an efficient alternative to traditional fine-tuning in scenarios with limited data~\citep{min2023recent, szép2024practicalguidefinetuninglanguage}. Such prompting strategies are pivotal in models like the GPT series, emphasizing instructions and demonstrations, and in template-based learning, where examples are integrated into natural text formats~\citep{min2023recent}. Furthermore, zero- and few-shot learning have emerged as powerful strategies in prompt engineering. Zero-shot learning involves crafting prompts that enable the model to generate useful responses without prior examples, relying solely on pre-trained knowledge. In contrast, few-shot learning involves providing a few examples within the prompt, guiding the model to understand the task context and desired response format. This technique has shown remarkable effectiveness in adapting models to new tasks with minimal examples. Several works have recently focused on few-shot prompting techniques, particularly to improve the reasoning capabilities of LLMs, such as the Chain-of-Thought~\citep{wei2022chain}, Tree-of-Thoughts~\citep{yao2023tree}, or Graph-of-Thoughts prompting~\citep{besta2023graph}.

Another technique to significantly improve the quality of user interactions and enhance user experiences during interactions with LLMs is prompt optimization through reformulation~\citep{wang2023promptagent}. By automating the process of prompt refinement, such strategies ensure higher accuracy and contextual appropriateness of the responses of LLMs. This improvement directly contributes to the field of HCI by enabling more intuitive and effective communication between users and AI systems. One fundamental approach in this context is templating, where prompts are structured in a consistent format to elicit specific types of responses. This method uses careful wording and phrasing to guide the model towards the desired output, thereby enhancing the predictability and reliability of interactions, which are key goals in HCI.

Additionally, using context reiteration and clarification in prompts has proven effective. By repeating or rephrasing key parts of the prompt, the clarity and focus of the model's responses can be significantly enhanced~\citep{dang2022prompt}. Although this method partially involves intent recognition, our work emphasizes precise intent recognition, followed by reformulation rather than relying on keyword detection, which may fail to recognize the actual semantics and intent of the prompt. The proposed technique aligns with HCI principles by ensuring that the system's responses are more aligned with user expectations and needs, reducing misunderstandings. Lastly, explicit instructions or constraints within prompts have been employed to direct the model's responses more precisely, ensuring adherence to specific guidelines or objectives. This approach supports the creation of user-centric AI systems that are adaptable to diverse interaction scenarios.

\section{Towards a Comprehensive and Fine-grained Taxonomy of Intents in User Prompts}
\label{sec:taxonomy}

Taxonomies have historically played an important role in information retrieval and knowledge representation, especially for the systematic organization of large volumes of information, providing a standardized classification across various systems~\citep{medelyan2013automatic}. A clear hierarchical structure can simplify search and discovery and facilitate understanding of contextual relationships between concepts. Along similar lines, in this section, we introduce a comprehensive and fine-grained taxonomy of user intents to address current limitations of user interactions with LLMs, specifically the user dissatisfaction~\citep{kim2023understanding} when LLMs fail to accurately interpret user intentions. 
  
\subsection{Important Characteristics for an Effective Intent Taxonomy}

An effective taxonomy for guiding intent-based interactions with LLMs must have certain key features~\citep{shah2023using}. First, it should provide a \textbf{comprehensive intent coverage} that encompasses a wide range of user intents, from factual queries to personal and creative interactions. To achieve this, we have carefully analyzed interdisciplinary literature on user intent and manually assembled a comprehensive and detailed intent taxonomy in the Appendix (see Table~\ref{tab:taxonomy_detailed}). Such comprehensive coverage is critical to accurately identify the intent behind user queries and enable LLMs to provide relevant and targeted responses. Second, the taxonomy should provide a \textbf{clear, precise, and consistent} categorization of intents. This clarity and precision are critical to extracting the most appropriate and accurate responses from LLMs. To achieve this, we have manually grouped the intents into consistent and coherent intent categories, as depicted in Table~\ref{tab:taxonomy_detailed}. Finally, it is essential to have a taxonomy that is \textbf{versatile and applicable across various applications and use-cases}, ranging from technical and educational contexts to personal and artistic interactions. This versatility is key to expanding the user-centered utility of LLMs in different domains. Since we  have used the insights from interdisciplinary literature, the required versatility is a key feature of the proposed intent taxonomy.

\subsection{A Fine-Grained Intent Taxonomy}


\begin{table}[t]
\caption{Taxonomy of Intent Types for User Prompts to MFMs.}
\centering
\small 
\begin{tabular}{>{\bfseries}m{1.7cm} m{13cm}}
\toprule
Intent Type & \textbf{Fine-granular Intent} \\
\midrule
Informational Intent & 
    \begin{itemize}
        \item \textbf{Factual Queries:} Requests for specific facts or data
        
        \item \textbf{Explanatory Inquiries:} Explanations or clarifications about concepts, events, phenomena
        
        \item \textbf{Tutorial Requests:} Step-by-step instructions or guidance
    \end{itemize}\\
\hline
Problem-Solving Intent & 
    \begin{itemize}
        \item \textbf{Troubleshooting Assistance:} Diagnose and resolve issues or problems
        
        \item \textbf{Decision Support:} Assistance in decision-making through insights, comparisons, evaluations
        
        \item \textbf{Planning and Organization:} Aid in planning events, organizing tasks, or managing projects
    \end{itemize}\\
\hline
Creative Intent & 
    \begin{itemize}
        \item \textbf{Idea Generation:} Inspiration or ideas for creative projects
        
        \item \textbf{Content Creation:} Help in writing, visually representing, or designing original content
        
        \item \textbf{Artistic Exploration:} Exploration of artistic styles, techniques, historical art contexts
    \end{itemize}\\
\hline
Educational Intent &
    \begin{itemize}
        \item \textbf{Learning Support:} Assistance with understanding educational material or concepts
        
        \item \textbf{Skill Development:} Guidance on developing specific skills or competencies
        
        \item \textbf{Curricular Planning:} Help in designing or choosing educational curricula or courses
    \end{itemize}\\
\hline
Personal Interaction Intent & 
    \begin{itemize}
        \item \textbf{Conversational Engagement:} Dialogue for entertainment, companionship, interaction
        
        \item \textbf{Personal Advice:} Advice on personal matters or life decisions
        
        \item \textbf{Reflection and Insight:} Help in self-reflection, personal growth, or to gain insights into certain behavior or thoughts
    \end{itemize}\\
\hline
Technical and Professional Intent & 
    \begin{itemize}
        \item \textbf{Technical Guidance:} Assistance with technical tasks, coding, or problem-solving in a professional context
        
        \item \textbf{Business and Career Advice:} Guidance on business, career choices, or professional development
        
        \item \textbf{Industry-Specific Inquiries:} Requests for information or assistance specific to certain industries or professional fields
    \end{itemize}\\
\hline
Transactional Intent & 
    \begin{itemize}
        \item \textbf{Service Utilization:} Requests to use specific functionalities of the model (e.g., language translation, summarization)
        
        \item \textbf{Data Processing:} Help in processing, analyzing, or visualizing data
        
        \item \textbf{Task Automation:} Inquiries about automating tasks or workflows
    \end{itemize}\\
\hline
Ethical and Philosophical Intent & 
    \begin{itemize}
        \item \textbf{Moral and Ethical Queries:} Questions about ethical dilemmas, moral principles, or philosophical theories
        
        \item \textbf{Societal and Cultural Inquiry:} Exploring societal, cultural, or historical topics
        
        \item \textbf{Existential Questions:} Delving into existential themes or abstract philosophical questions 
    \end{itemize}\\
\bottomrule
\end{tabular}
\label{tab:taxonomy}
\end{table}


User interactions with LLMs are multifaceted, reflecting a wide spectrum of needs and purposes. To systematically categorize the different intents behind such interactions, we drew from recent research on user-LLM engagement~\citep{shah2023using} and emerging trends in human-AI interaction~\citep{MediumChatGPTStats, SparktoroChatGPT} (see Table~\ref{tab:taxonomy_detailed}). For completeness, our approach involved an initial analysis of common web search types -- namely, navigational, informational, and transactional queries~\citep{broder2002taxonomy} -- further refined by examining granular subcategories~\citep{jansen2008determining}. Additionally, \cite{kellar2007field} observed users interacting for various information-seeking tasks, such as fact-finding, information gathering, browsing, and transactions, introducing further subcategories in this intent. Besides navigational and informational intents, \cite{rose2004understanding} also examined resource-oriented intents, like interaction, entertainment, downloading, and obtaining resources. They explored user goals such as seeking for specific information, open-ended questions, undirected information about a topic, advice, locating services, and products~\citep{rose2004understanding}. In another study, \cite{ashkan2009classifying} focused on classifying commercial intents to personalize search outcomes and boost user satisfaction, while \cite{yi2008deciphering} categorized mobile search queries by topics, employing an in-house taxonomy that included 821 subcategories organized under 23 top-level categories, such as entertainment, finances, government and politics, health and pharma, hobbies, life stages, news, reference, religion, science, sports, technology, and travel~\citep{yi2008deciphering}. 

By manually extracting all the intents we encountered in the above literature and categorizing them according to the corresponding articles, we derived the very detailed taxonomy from depicted in Table~\ref{tab:taxonomy_detailed}. However, to make the taxonomy manageable for our study and of practical use, we transformed the extensive and detailed taxonomy into a more concise and practical set of fine-grained intents by:
\begin{itemize}
    \item Grouping similar sub-intents to further eliminate potential redundancy.
    \item Selecting key representative intents that are most relevant to our study.
    \item Designing specific reformulation templates for these selected intents to guide the LLMs effectively.
    \item Ensuring the templates are aligned with the characteristics of effective prompt reformulation outlined above.
\end{itemize}

This approach allowed us to maintain the essence and coverage of the original taxonomy while making it operationally feasible for the user study presented in this paper. The final taxonomy strikes a balance between comprehensiveness and practicality, and enables the investigation of the impact of intent-based prompt reformulation on user satisfaction with LLM responses. 


\section{User Study: Analysis of Intent Recognition and Intent-Based Prompt Reformulation}
\label{sec:study_description}

To evaluate the accuracy of intent recognition and the effect of intent-based prompt reformulation on user satisfaction, we designed and conducted a three-phase user study as described in the following subsections. More details can also be found in the Appendix.

\subsection{User Study Phase 1: Quality of Intent Recognition by State-of-the-Art LLMs} 

The first phase of the user study was devoted to assessing the accuracy of intent recognition by comparing two different LLMs, namely GPT-3.5 Turbo (i.e., gpt-3.5-turbo-1106) and GPT-4 Turbo (i.e., gpt-4-1106-preview), in a between-subjects design fashion. For the sake of readability, in the remainder of this paper, we refer to these models as GPT-3.5 and GPT-4, respectively. We selected these models as they currently represent the most widely used LLMs available, both in their free versions and paid subscriptions, across different application domains. This comparison aimed to uncover how advances in model architecture might impact user experience in real-world scenarios. In this part of the user study, the participants were introduced to conversational contexts extracted from a publicly accessible dataset~\citep{lhoest2021datasets} and were asked to continue these dialogues (Fig.~\ref{img:question}). The exact elements selected from this dataset are available on our GitHub repository in the \textit{chat\textunderscore history.json} file. The primary purpose of these conversation starters (i.e., conversational contexts) was twofold: to restrict questions to the given contexts and to guide and expand the range of intent categories emerging during the study. The underlying LLM analyzed the user's prompt to recognize their intent based on the taxonomy introduced in Table~\ref{tab:taxonomy}. The participants were then asked, whether they agreed with the intent category identified by the LLM (Fig.~\ref{img:class}). 
In cases where participants disagreed with the detected intent, they were asked to select a more suitable category from our intent taxonomy, as detailed in Table~\ref{tab:taxonomy} (Fig.~\ref{img:not_agree}). This process was repeated ten times for various conversational contexts. 

\subsection{User Study Phase 2: Effect of Intent-based Prompt Reformulation on User Satisfaction} 

The primary objective of the second phase of our user study was to assess how well the LLMs' responses, both to the original and reformulated user prompts, align with user preferences. Following the first phase, which assesses the intent recognition process, each participant was presented with two distinct responses to their prompt in the second phase of the study (Fig.~\ref{img:answer}). 
The underlying LLM generated one of the responses directly from the original user prompt, while the other response was generated based on the reformulated user prompt aligning with the correctly identified intent category, as presented in Table~\ref{table:reformulation-templates}. To maintain objectivity in participant responses, the participants were blind to the answer options, and the presentation order of responses was randomized. Furthermore, to mitigate bias associated with the length of answers, both responses were kept to approximately the same length by prompt modification: "Please answer the following question in a maximum 2500 characters: \textit{-user question-}". Participants were then asked to choose the more suitable or preferable response. This phase of the user study provided insights into user preferences regarding AI-generated responses, both to original and reformulated prompts. A detailed description of the prompt reformulation based on intent classification is provided in the following subsections.

\begin{figure}[t]
\includegraphics[width =1\textwidth]{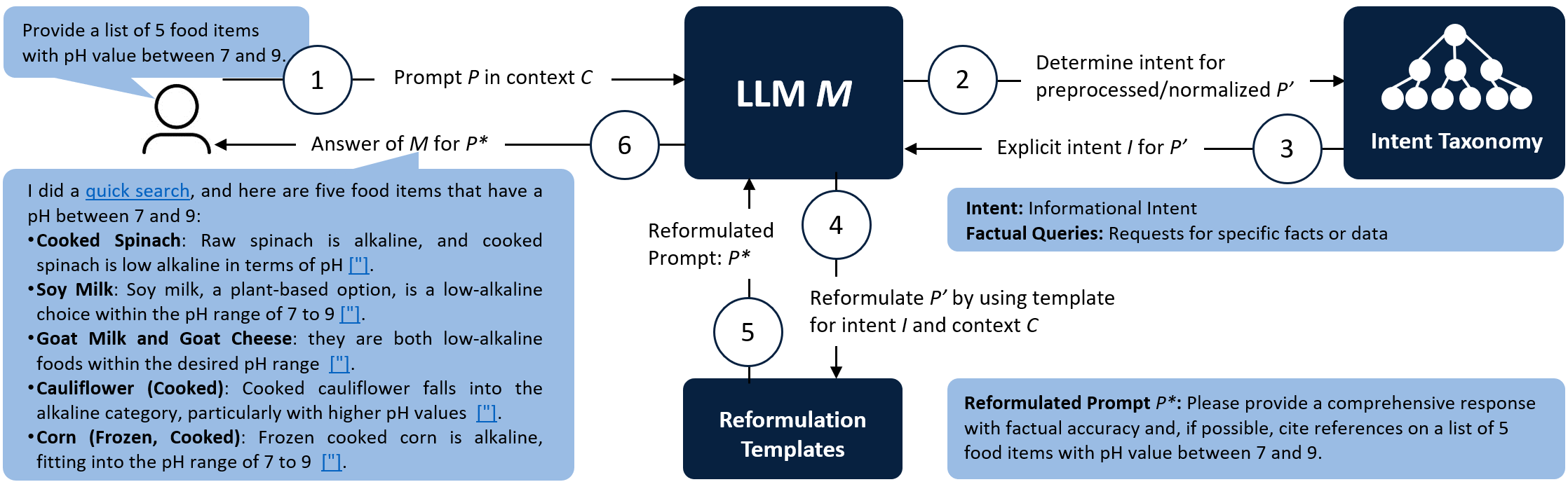}
\caption{Overview of the proposed prompt reformulation framework, visual summary of Algorithm~\ref{alg:procedure}.}
\label{img:approach}
\end{figure}

\subsubsection{Prompt Reformulation through Intent Classification} \label{sec:prompt_reformulation}
Advanced generative AI systems are trained on vast text corpora and can generate responses across various topics and scenarios. Hence, the main challenge is not including more knowledge into such systems, but rather precisely extracting the most pertinent and accurate responses to user prompts. Recent research on prompt reformulation often argues that conventional direct querying methods often fall short of leveraging the full potential of LLMs, leading to responses that, while correct, may not fully align with the user's intent or context. Thus, the intent-based reformulation strategy for our user study aims to investigate the link between the user's intent, the conversational context, and the available LLM knowledge. Figure~\ref{img:approach} and Algorithm~\ref{alg:procedure} present our prompt reformulation strategy.

As depicted in Algorithm~\ref{alg:procedure}, the framework begins by allowing the LLM ($M$) to preprocess the user's prompt ($P$) to ensure linguistic correctness, followed by the intent classification of the refined prompt ($P'$) using $M$. This process involves asking $M$ to categorize $P'$ based on predefined intent types from Table~\ref{tab:taxonomy}. The next step is to retrieve a concise summary of the conversational context ($C$) from $M$. Subsequently, a suitable template ($T$) for the identified intent is selected from a set of predefined templates from Table~\ref{table:reformulation-templates}. This template is then tailored to incorporate the conversational context $C$, forming a new, context-enhanced template ($T'$). The final step involves reformulating the user prompt using $T'$, resulting in a reformulated prompt that is then used by the LLM $M$ to generate a response. This approach ensures that the user's original intent is captured more accurately in a realistic, contextually relevant manner.

\begin{algorithm}[h]
\footnotesize
\Proc{ReformulateUserPrompt (\textit{P})}{
    $P' \gets \text{PreprocessPrompt}(M,P)$ \tcp*[r]{Try to express $P$ in correct English, in case of linguistic issues}
    $I \gets \text{ClassifyIntent}(M, P')$ \tcp*[r]{Ask $M$ to classify $P'$ according to one of the intent types in Table~\ref{tab:taxonomy}}
    $C \gets \text{AnalyzeGetContext}(M)$ \tcp*[r]{Get from $M$ a succinct summary of the conversational context $C$}
    $T \gets \text{SelectTemplate}(I)$ \tcp*[r]{Returns template $T$ from Table~\ref{table:reformulation-templates} for $I$}
    $T' \gets \text{IncorporateContext}(C, T)$ \tcp*[r]{I.e., a new template of the form [``In the context" \& C \& T] is created}
    $P^* \gets \text{ReformulatePrompt}(T', P')$ \\
    \Return \text{GenerateResponse}($M, P^*$)
}
\caption{User Prompt Reformulation Framework (LLM $M$, Prompt $P$)}
\label{alg:procedure}
\end{algorithm}

\subsubsection{The Prompt Reformulation Templates}
For the proposed templates, we aim to generate prompt prefixes satisfying the following desiderata:

\begin{itemize}

    \item \textbf{Clear intent representation:} The template should be designed to interpret and respond to the expressed intent, not just the literal prompt.
    
    \item  \textbf{Clear and precise goal description:} The template should clearly and precisely specify the expected response type, level, and detail. The more specific the template, the more targeted the model's response will be. 

    \item \textbf{Flexibility:} The template should be adaptable to a range of prompts of the same intent without requiring substantial modifications for different types of information.

    \item \textbf{Reference indication for correctness support:} If needed, especially for informational queries, the template should include a request for references to ensure the credibility of the response~\footnote{This last point is optional because the model's ability to provide actual references might depend on its training and access to retrieval systems on external data.}. 
    
\end{itemize}

Specifically, for each intent type and each corresponding request type from the proposed taxonomy (Table~\ref{tab:taxonomy}), we asked ChatGPT (GPT-4) to provide the most appropriate template according to the above desiderata that could be used as a prompt prefix for the specific query type. In addition, as a post-hoc test to assure the quality of the generated templates, we asked ChatGPT for each template whether it adequately reflects the above desiderata and to improve the template accordingly in case one or more desiderata can be better reflected. Interestingly, an improvement was suggested for some templates. For example, for the intent of `learning support', ChatGPT suggested the following adjustment: 

\begin{center}
\begin{lrbox}{\CBox}
\begin{minipage}{0.8\linewidth}
\textBF{Original:} [Provide comprehensive educational support and resources for a deeper understanding of] \\
\textBF{Assessment:} Clear and flexible, but could specify the nature of educational support. \\
\textBF{Improvement:} [Offer educational support through explanations, examples, or resources for a comprehensive understanding of]
\end{minipage}
\end{lrbox}

\begin{Sbox}
\begin{minipage}{0.8\linewidth}
\textBF{Original:} [Provide comprehensive educational support and resources for a deeper understanding of] \\
\textBF{Assessment:} Clear and flexible, but could specify the nature of educational support. \\
\textBF{Improvement:} [Offer educational support through explanations, examples, or resources for a comprehensive understanding of]
\end{minipage}
\end{Sbox}
\fcolorbox{black}{mygrey}{\TheSbox} 
\end{center}

For other intents, no adjustment was needed. For instance, for `skill development', the answer of ChatGPT was: 

\begin{center}
\begin{lrbox}{\CBox}
\begin{minipage}{0.8\linewidth}
\textBF{Original:} [Offer detailed guidance and practical tips for skill enhancement in] \\
\textBF{Assessment:} Specific and adaptable to different skills. \\
\textBF{Improvement:} [None required.]
\end{minipage}
\end{lrbox}

\begin{Sbox}
\begin{minipage}{0.8\linewidth}
\textBF{Original:} [Offer detailed guidance and practical tips for skill enhancement in] \\
\textBF{Assessment:} Specific and adaptable to different skills. \\
\textBF{Improvement:} [None required.]
\end{minipage}
\end{Sbox}
\fcolorbox{black}{mygrey}{\TheSbox} 
\end{center}

The final suggestions were considered the most appropriate templates to express the related intents. Table~\ref{table:reformulation-templates} gives an overview of all templates generated in this way and employed in the second phase of the user study. 

\begin{table*}[!ht]
\centering
{\small
\begin{tabular}{ p{4cm} p{12cm} }
\toprule
\textbf{Detailed Prompt Intent} & \textbf{Reformulation Template} \\
\midrule
\multirow{2}{*}{\textbf{Factual Queries}} & [Provide a comprehensive response with factual accuracy and, if possible, cite references on ...] \\
\multirow{2}{*}{\textbf{Explanatory Inquiries}} & [Elucidate the concept of ..., ensuring a detailed explanation of key aspects and implications, adaptable to various topics in ...] \\
\textbf{Tutorial Requests} & [Provide a step-by-step tutorial or instruction on how to effectively ...] \\
\hline
\textbf{Troubleshooting Assistance} & [Assist in identifying and solving the problem related to ..., considering possible solutions for ...] \\
\multirow{2}{*}{\textbf{Decision Support}} & [Offer an evaluation and comparison of advantages and drawbacks for various options regarding ...] \\
\textbf{Planning and Organization} & [Outline a structured plan with key steps and considerations for efficiently organizing ...] \\
\hline
\textbf{Idea Generation} & [Suggest innovative ideas or creative approaches, adaptable to different contexts for ...] \\
\textbf{Content Creation} & [Assist in creating engaging content, such as articles, videos, etc., focused on ...] \\
\textbf{Artistic Exploration} & [Explore and discuss various artistic approaches and styles suitable for ...] \\
\hline
\multirow{2}{*}{\textbf{Learning Support}} & [Offer educational support through explanations, examples, or resources for a comprehensive understanding of ...] \\
\textbf{Skill Development} & [Offer detailed guidance and practical tips for skill enhancement in ...] \\
\multirow{2}{*}{\textbf{Curricular Planning}} & [Assist in selecting or developing a curriculum, focusing on subjects, levels, and educational goals for ...] \\
\hline
\textbf{Conversational Engagement} & [Engage in an interactive and thoughtful conversation about ...] \\
\textbf{Personal Advice} & [Provide thoughtful and considerate personal advice regarding ...] \\
\multirow{2}{*}{\textbf{Reflection and Insight}} & [Encourage self-reflection and offer insights, adaptable to various personal or professional topics on ...] \\
\hline
\textbf{Technical Guidance} & [Offer in-depth technical guidance and support for issues related to ...] \\
\textbf{Business and Career Advice} & [Provide strategic guidance and advice, adaptable to various business and career paths on ...] \\
\textbf{Industry-Specific Inquiries} & [Present detailed and industry-specific insights and information about ...]  \\
\hline
\textbf{Service Utilization} & [Instruct me in a detailed way and step-by-step on how to use ...] \\
\textbf{Data Processing} & [Assist me in processing and analyzing ...] \\
\textbf{Task Automation} & [Provide specific guidance on automating ...] \\
\hline
\textbf{Moral and Ethical Queries} & [Engage in a thoughtful examination and discussion of the moral and ethical aspects of ...] \\
\textbf{Societal and Cultural Inquiry} & [Investigate and discuss the societal and cultural dimensions of ...] \\
\textbf{Existential Questions} & [Delve into and discuss philosophical perspectives and viewpoints on ...] \\
\bottomrule
\end{tabular}
}
\caption{Reformulation Templates for Detailed Intent Types.}
\label{table:reformulation-templates}
\end{table*}

\subsection{User Study Phase 3: User's Understanding of Prompt Reformulation}
In the last phase of the user study, we collected participants' demographic data and focused on their understanding of the prompt reformulation concept. In addition, we provided three prompt reformulation examples and asked them to reformulate simple prompts. To evaluate participants' understanding of prompt reformulation in the context of content creation, we asked them to rate their likelihood of using the provided prompt reformulation templates on a 5-point Likert scale, where 1 indicated `extremely unlikely' and 5 indicated `extremely likely'. This stage of the user study aimed to assess participants' willingness to utilize the templates after they were presented alongside a sample prompt on the same page.

\subsection{Participant Recruitment}
In our user study utilizing GPT-3.5 and GPT-4, we engaged two distinct groups of participants through the Prolific platform. This choice was made considering Prolific's reputation for yielding higher quality data~\citep{Peer_etal_2022}. We ensured a gender-balanced sample, with participants 18 years old or older, and fluent in English. Once participants were recruited on Prolific, they were forwarded to Qualtrics, where our user study was implemented, and upon the completion of the study, they were redirected to Prolific for compensation. Participants were compensated according to \texteuro15/hour rate for participation in our studies. Because of the difference in the necessary time for the API calls to the two language models, the time frames to complete the study ranged between 40 and 50 minutes, with GPT-3.5 as the underlying LLM and GPT-4, respectively. The data was collected anonymously, and each participant provided digital informed consent before starting the study. Participants had the option to withdraw their consent or leave the study at any point, without the need for any further explanation. In total, we recruited $n_{GPT-3.5} = 124$ $(M_{age} = 30.9, SD_{age} = 11.2)$ for the GPT-3.5 and $n_{GPT-4} = 116$ $(M_{age} = 28.3, SD_{age} = 8.0)$ for the GPT-4 study, respectively. We filtered the participants out if there was a reported age inconsistency between Prolific and Qualtrics and a mismatch in gender, which led us to $n_{GPT-3.5} = 120$ and $n_{GPT-4} = 114$ participants, respectively. Upholding ethical standards, we collected the anonymized user data based on explicit user consent. We obtained written consent from each participant, covering the storage, usage, and sharing of their conversational data for research purposes.

We also excluded participants who finished the study in an exceptionally short time. Considering the API response time for GPT-4 was nearly double that of GPT-3.5, and following an analysis of outliers based on the total duration of the study, we established time thresholds of 20 minutes for GPT-3.5 and 35 minutes for GPT-4. These thresholds also align with our pilot testing, where we determined that the ideal time range for each main page, corresponding to the models, is approximately between 1 and 2 minutes, leading to an expected time range between 25 and 45 minutes for the whole study per participant. If the participants were to spend less time within this interval, they might not have spent sufficient time to fully read and comprehend the GPT answers, making their choice less reliable. Three additional users had to be excluded from further analysis based on their answer quality, namely giving at least four inputs with only the space character or asking the same questions such as `Is that true?' throughout the survey. After these relatively conservative exclusions, there were $n_{GPT-3.5} = 116$ and $n_{GPT-4} = 95$ participants for GPT-3.5 and GPT-4, respectively.

\section{Results and Discussion}
\label{sec:results}
This section evaluates the intent recognition capabilities of GPT-3.5 and GPT-4, analyzing their accuracy and the impact of prompt reformulation on user satisfaction. We explore both quantitative results and qualitative user feedback, discuss our key findings, and highlight future research directions.

\subsection{Quality of Intent Recognition by State-of-the-Art LLMs}
As we asked the participants to subjectively evaluate the intent recognition results provided by the GPT models, we first assessed whether participants' self-reported intent categories align with those provided by the GPT models. This is necessary as the intent recognition process is essential for the subsequent prompt reformulation. To this end, we employ descriptive statistics and accuracy measures and report confusion matrices for the performance of intent recognition of both GPT-3.5 and GPT-4.

\begin{table*}[htb]
    \centering
    \begin{tabular}{>{\centering\arraybackslash}p{11mm}|>{\centering\arraybackslash}p{8mm} >{\centering\arraybackslash}p{8mm} >{\centering\arraybackslash}p{8mm} >{\centering\arraybackslash}p{8mm} >{\centering\arraybackslash}p{8mm} >{\centering\arraybackslash}p{8mm} >{\centering\arraybackslash}p{8mm} >{\centering\arraybackslash}p{8mm} >{\centering\arraybackslash}p{8mm} >{\centering\arraybackslash}p{8mm} >{\centering\arraybackslash}p{8mm} >{\centering\arraybackslash}p{8mm}}
   & \rotatebox{90}{Factual Queries} & \rotatebox{90}{\makecell{Explanatory\\Inquiries}} & \rotatebox{90}{\makecell{Tutorial\\Requests}} & \rotatebox{90}{\makecell{Troubleshooting\\Assistance}} & \rotatebox{90}{\makecell{Decision\\Support}} & \rotatebox{90}{\makecell{Planning and\\Organization}} & \rotatebox{90}{\makecell{Idea\\Generation}} & \rotatebox{90}{\makecell{Content\\Creation}} & \rotatebox{90}{\makecell{Artistic\\Exploration}} & \rotatebox{90}{\makecell{Learning\\Support}} & \rotatebox{90}{\makecell{Skill\\Development}} & \rotatebox{90}{\makecell{Curricular\\Planning}} \\
    GPT-3.5 & 75.50 & 66.34 & 59.26 & 57.14 & 97.56 & \textbf{91.67} & 84.00 & 36.67 & 0.00 & \textbf{36.84} &
       \textbf{66.67} & \textbf{50.00}  \\
        GPT-4 & \textbf{96.02} & \textbf{89.50} & \textbf{98.39} & \textbf{80.95} & \textbf{98.63} &
       77.78 & \textbf{85.71} & \textbf{90.48} & \textbf{100} & 0.00 & 55.56 & 0.00  \\
    \end{tabular}

    \begin{tabular}{>{\centering\arraybackslash}p{11mm}|>{\centering\arraybackslash}p{8mm} >{\centering\arraybackslash}p{8mm} >{\centering\arraybackslash}p{8mm} >{\centering\arraybackslash}p{8mm} >{\centering\arraybackslash}p{8mm} >{\centering\arraybackslash}p{8mm} >{\centering\arraybackslash}p{8mm} >{\centering\arraybackslash}p{8mm} >{\centering\arraybackslash}p{8mm} >{\centering\arraybackslash}p{8mm} >{\centering\arraybackslash}p{8mm} >{\centering\arraybackslash}p{8mm}}
        & \rotatebox{90}{\makecell{Conversational\\Engagement}} & 
  \rotatebox{90}{\makecell{Personal\\Advice}} & 
  \rotatebox{90}{\makecell{Reflection\\and Insight}} & 
  \rotatebox{90}{\makecell{Technical\\Guidance}} & 
  \rotatebox{90}{\makecell{Business and\\Career Advice}} & 
  \rotatebox{90}{\makecell{Industry-Specific\\Inquiries}} & 
  \rotatebox{90}{\makecell{Service\\Utilization}} & 
  \rotatebox{90}{\makecell{Data\\Processing}} & 
  \rotatebox{90}{\makecell{Task\\Automation}} & 
  \rotatebox{90}{\makecell{Moral and\\Ethical Queries}} & 
  \rotatebox{90}{\makecell{Societal and\\Cultural Inquiry}} & 
  \rotatebox{90}{\makecell{Existential\\Questions}} \\
        GPT-3.5 & 71.43 & \textbf{93.68} & 82.61 & \textbf{80.00} & 0.00 & \textbf{52.38} & 80.00 & \textbf{66.67} & 90.16 & 90.43 & 46.15 & 0.00 \\
        GPT-4 & \textbf{92.86} & 88.24 & \textbf{100} & 70.00 & \textbf{66.67} & 25.00 & \textbf{90.91} & 0.00 & \textbf{100} & \textbf{95.92} & \textbf{81.25} & \textbf{77.27}     
    \end{tabular}
    \caption{Category-wise accuracy values for both models. Due to the low frequency of some unusual intent categories in the dataset, the recognition rate of 0.00\% is a rather conservative performance estimate for both models. (See Figure~\ref{fig:confmx} for details.)}
    \label{tab:classwise_acc}
\end{table*}

\begin{figure}[t]
	\centering
	\subfloat[Confusion matrix for GPT-3.5.]{\includegraphics[height=9cm, keepaspectratio]{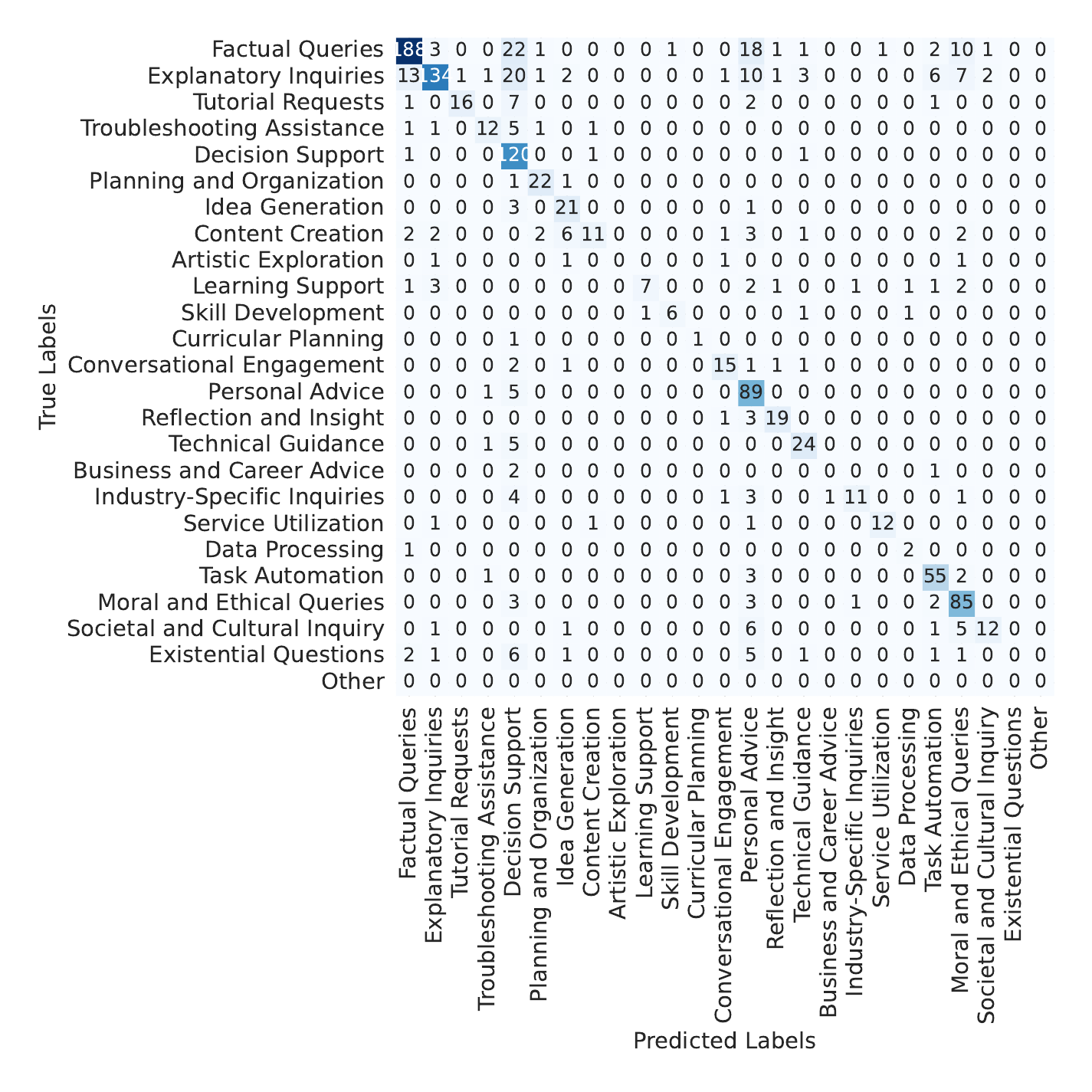}
    \label{fig:GPT-3.5 Turbo-cm}}
	\subfloat[Confusion matrix of GPT-4.]{\includegraphics[height=9cm, keepaspectratio]{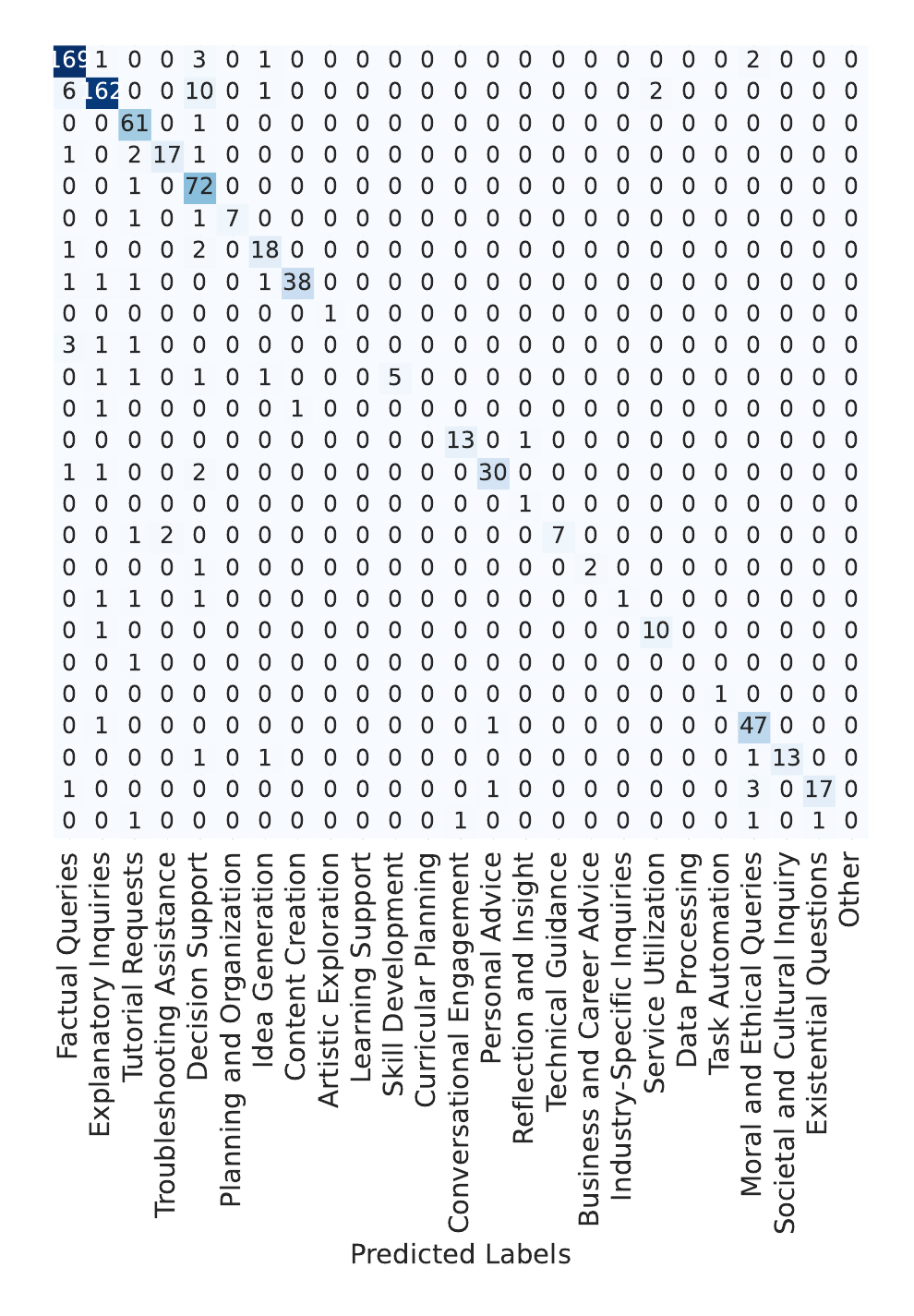}
	\label{fig:gpt4-cm}}
	\hspace{10mm}
	\caption{Details about the exact intent classification, represented in confusion matrices.}
    \label{fig:confmx}
\end{figure}

More specifically, based on each conversational context and the subsequent user query, we asked the respective LLM model to categorize the user intent into the predefined 24 fine-grained intent categories from Table~\ref{tab:taxonomy}. With this approach, we achieved accuracy values of 75.28\% and 89.64\% for GPT-3.5 and GPT-4, respectively. In addition, $F_1$ scores of 74.28\% and 88.84\% were obtained using GPT-3.5 and GPT-4, respectively (see also Table~\ref{tab:gpt-metric-comp}). Relevant confusion matrices are depicted in Figure~\ref{fig:confmx}. According to these results, while GPT-3.5 struggled with classifying factual queries and explanatory inquiries, GPT-4 achieved superior performance by 20.5 and 23.17 percent points in accuracy for the mentioned categories, respectively. The increase of the model performance in the content creation category is also significant, as GPT-4 reached an increased accuracy of 53.81 percent points. However, both models struggled with recognizing the `learning support' intent. Considering all categories, GPT-4 achieved better accuracy in 16 out of 24 intent categories (Table~\ref{tab:classwise_acc}). 

\begin{table}[ht]
\centering
\begin{tabular}{ccc}
     \toprule
     Model & Accuracy [\%] & $F_1$ score [\%]\\
\midrule
GPT-3.5 Turbo & 75.28 & 74.28  \\
GPT-4 Turbo & \textbf{89.64} & \textbf{88.84}   \\ 
\bottomrule
\end{tabular}
\caption{Intent classification accuracy and $F_1$ score values.}
\label{tab:gpt-metric-comp}
\end{table}

\subsection{Effect of Intent-based Prompt Reformulation on User Satisfaction} 

In the second phase of the user study, we analyzed user preferences for responses generated by GPT-3.5 and GPT-4 with and without leveraging prompt reformulation aligned with the previously determined intent category. The collected dataset is available on~\href{https://github.com/ConcealedIDentity/UserIntentStudy}{GitHub}.

We conducted separate analyses for each of the 24 detailed intent categories to determine the number of participants who favored either the original GPT responses or the GPT responses to reformulated user prompts for both models independently. To this end, we utilize paired samples t-tests with a confidence level of 0.95, i.e., $\alpha=0.05$, separately for GPT-3.5 and GPT-4. Furthermore, we employed paired samples t-tests, maintaining the same alpha level, to assess if GPT-3.5 and GPT-4 show comparable performance.  

Our analysis only included instances where GPT correctly identified the user intent in the first phase, as incorrect intent classification could result in the applied template altering the direction of the reformulated answer. Furthermore, we implemented a preliminary filtering step on the collected samples to ensure a fair and consistent analysis. This process involved comparing the length of GPT's responses to the original user prompts with those to the reformulated prompts. More specifically, We only included responses in our evaluation where the length difference between the two sets of answers was within a 10\% margin in terms of character count. This approach was adopted to mitigate any potential bias that might arise from variations in answer length. 

\begin{figure}[t]
	\centering
	\subfloat[Data distribution with GPT-3.5 model.]{\includegraphics[height=6.1cm, keepaspectratio]{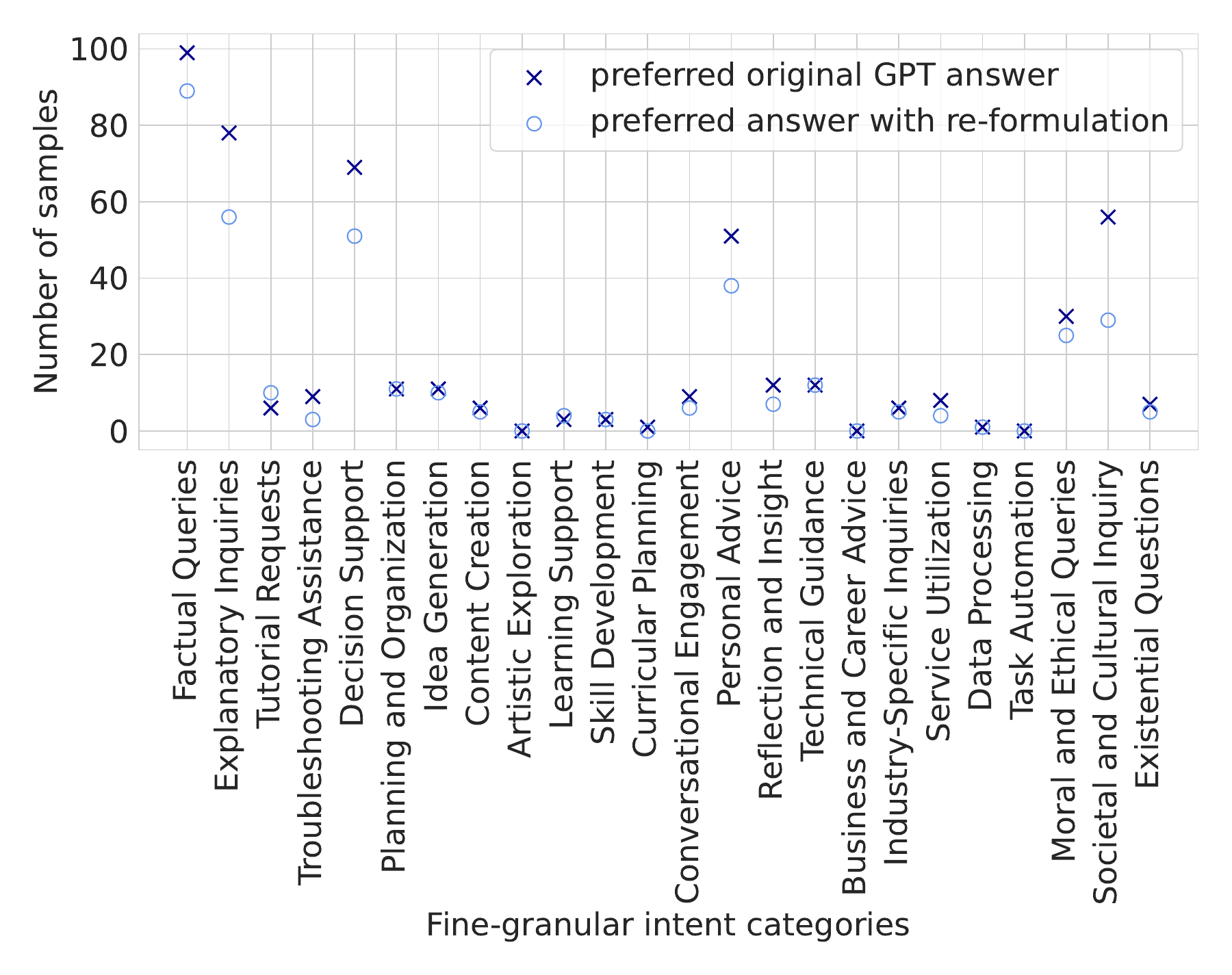}
    \label{fig:GPT-3.5 Turbo}}
	\subfloat[Data distribution with GPT 4 Turbo model.]{\includegraphics[height=6.1cm, keepaspectratio]{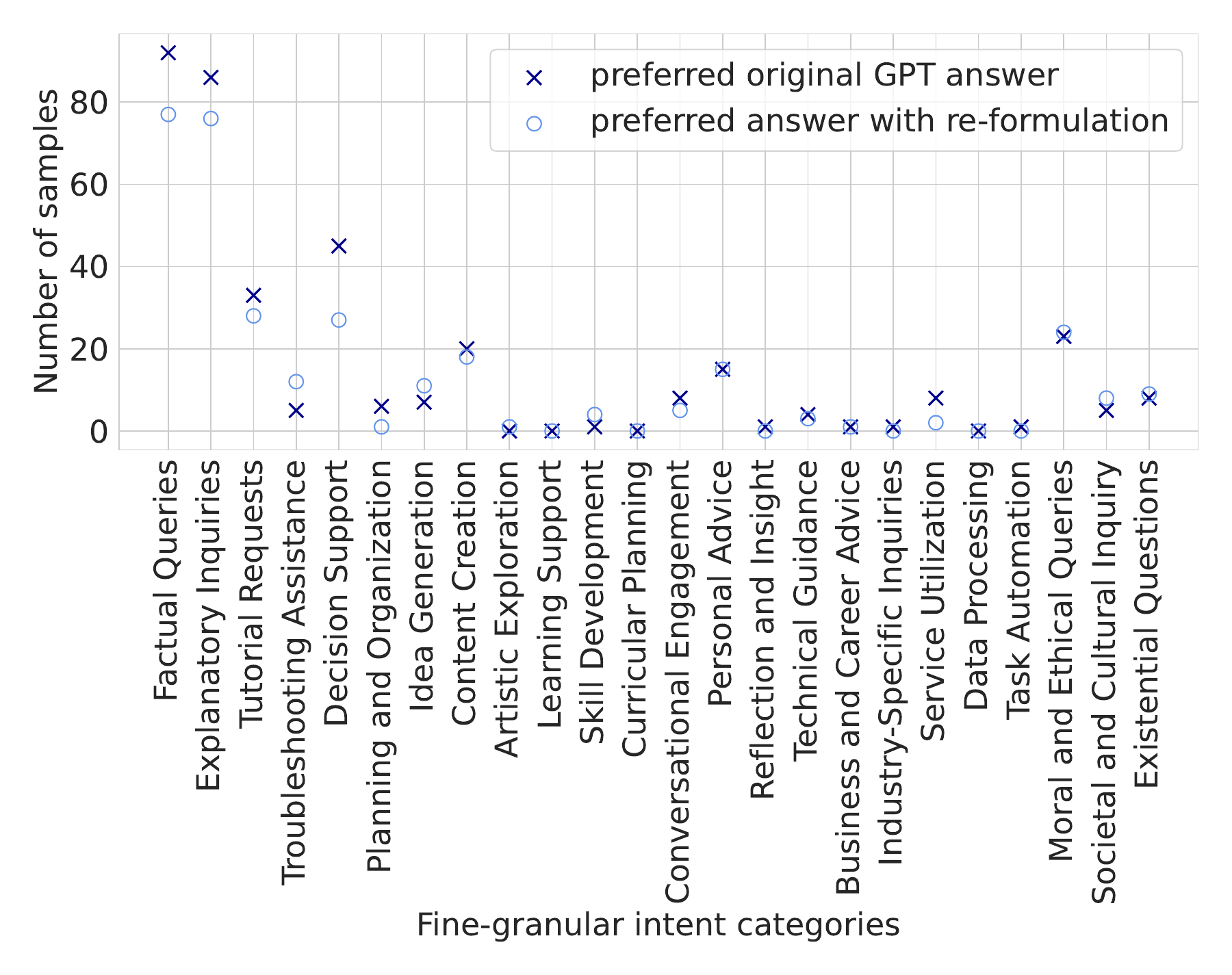}
	\label{fig:gpt4}}
	\hspace{10mm}
	\caption{Data distribution corresponding to the fine-granular intent categories.}
    \label{fig:base_filter}
\end{figure}

By analyzing the answer preferences of the users, we compared the number of samples category-wise (Figure~\ref{fig:base_filter}). Interestingly, participants preferred the answers to the reformulated prompts generated by GPT-4 over those by GPT-3.5. However, as indicated in Figure~\ref{fig:base_filter}, in both GPT-3.5 and GPT-4 cases, users generally favored the original model answers, with a preference rate of 56.61\% for GPT-3.5 and 53.5\% for GPT-4, respectively. For certain categories, users preferred the answer to the reformulated prompts. Specifically, in the context of GPT-3.5, this trend was observed in categories like `tutorial requests' and `learning support'. Meanwhile, with GPT-4, users favored the model answers to their reformulated prompts in areas such as `troubleshooting assistance', `idea generation', `skill development', `moral and ethical queries', and `societal and cultural inquiries', see also Table~\ref{tab:categories}. This observation underscores that the higher the advancement level of the GPT model, the better the quality of the answers to the reformulated prompts. Notably, in the `factual queries' category, our template is designed to prompt the model to include references in its responses. Upon examining the quantity of references in the responses to the reformulated prompts, GPT-4 demonstrated superior performance compared to GPT-3.5. However, it is noteworthy that only 25.53\% and 40.24\% of the responses incorporated references (see Table~\ref{tab:references}), either in citation or link format. This discrepancy in reference inclusion has significant implications for user-LLM interactions. It suggests that while advanced models like GPT-4 show promise in enriching responses with references, there is a clear need for further refinement to consistently meet user expectations for detailed and sourced information. In evaluating user preferences for the inclusion of references, we found that participants preferred the responses to reformulated queries produced by GPT-3.5 in 58.33\% of the cases. In contrast, users favored GPT-4 answers to reformulated prompts in only 35.29\% of instances, as depicted in Table~\ref{tab:user_pref_factual}. The following tables summarize these findings.

\begin{table}[htbp]
    \centering
    \caption{User Preference for Original vs.\ Reformulated Answers per Model}
    \label{tab:user_preference}
    \begin{tabular}{lcccc}
        \hline
        \textbf{GPT Model} & \textbf{Preferred Original Answers (\%)} & \textbf{Preferred Reformulated Answers (\%)} & \textbf{\textit{t-statistic}} & \textbf{\textit{p-value}}\\
        \hline
        GPT-3.5 & 56.61 & 43.39 & 2.9821 & 0.0067\\
        GPT-4   & 53.50 & 46.50 & 1.7250 & 0.0979\\
        \hline
    \end{tabular}
\end{table}

\begin{table}[htbp]
    \centering
    \caption{Categories Where Users Preferred Answers to Reformulated Prompts}
    \label{tab:categories}
    \begin{tabular}{lp{9cm}}
        \hline
        \textbf{GPT Model} & \textbf{Categories} \\
        \hline
        GPT-3.5 & Tutorial Requests, Learning Support \\
        GPT-4   & Troubleshooting Assistance, Idea Generation, Skill Development, Moral and Ethical Queries, Societal and Cultural Inquiries \\
        \hline
    \end{tabular}
\end{table}

\begin{table}[htbp]
    \centering
    \caption{Inclusion of References in Responses to Reformulated Prompts in the ``Factual Queries'' Category}
    \label{tab:references}
    \begin{tabular}{lc}
        \hline
        \textbf{GPT Model} & \textbf{Responses with References (\%)} \\
        \hline
        GPT-3.5 & 25.53 \\
        GPT-4   & 40.24 \\
        \hline
    \end{tabular}
\end{table}

\begin{table}[htbp]
    \centering
    \caption{User Preference Between GPT-3.5 and GPT-4 Responses to Reformulated Prompts in the ``Factual Queries'' Category}
    \label{tab:user_pref_factual}
    \begin{tabular}{lc}
        \hline
        \textbf{Preferred Model's Responses} & \textbf{Percentage (\%)} \\
        \hline
        GPT-3.5 & 58.33 \\
        GPT-4   & 35.29 \\
        \hline
    \end{tabular}
\end{table}

For a more comprehensive statistical analysis of the collected preference data, we further implemented paired t-tests (Table~\ref{tab:user_preference} and Table~\ref{tab:t-test-groups}). This approach was appropriate given that the aggregated preferences (and the differences in preference) conform to a normal distribution and meet other essential assumptions. According to the p-values obtained from this analysis, there is no substantial mean difference between the responses to original prompts and those to reformulated prompts for GPT-4, as well as in the comparison between the GPT-3.5 and GPT-4 datasets. However, a significant difference emerges when comparing the original responses to those generated in response to reformulated prompts for GPT-3.5.

\begin{table}[ht]
\centering
\begin{tabular}{ccc}
     \toprule
     & \textit{t}-statistic     & \textit{p}-value\\
\midrule 
GPT-3.5 Turbo  vs.  GPT-4 Turbo & 1.2506 & 0.2173 \\
\bottomrule
\end{tabular}
\caption{Paired t-test results of mean preferences of responses.}
\label{tab:t-test-groups}
\end{table}

\subsection{User's Understanding of Prompt Reformulation} 

After analyzing the two main parts of the study, we measured user satisfaction by asking the following two questions:
\begin{enumerate}
    \item In the Comparison Phase, the new answers were produced by the same large language model, but a prompt reformulation was used. Would you try reformulations if the templates were available to you?
    \item Here are some examples of how to use the reformulation templates:
        \begin{itemize}
            \item Factual Queries: Provide a comprehensive response with factual accuracy and, if possible, cite references on -\textit{your preferred topic}-
            \item Content Creation: Assist in creating engaging content, such as articles, videos, etc., focused on -\textit{your preferred topic}-
            \item Existential Questions: Delve into and discuss philosophical perspectives and viewpoints on -\textit{your preferred topic}-
        \end{itemize}
    Would you try out these reformulations to make your conversation with a large language model more effective?
\end{enumerate}

The average scores were 4.0 (Likely) for GPT-3.5 and 3.8 (Likely) for GPT-4, respectively, suggesting that users are open to learning how to formulate their prompts more effectively. Interestingly, when the users were asked to apply reformulation to a given prompt, 46.55\% and 54.74\% of users corresponding to GPT-3.5 and GPT-4, respectively, used the keywords from the correct template from the previously provided three different templates in their reformulated prompts (see also Table~\ref{tab:gpt-likert}). 

\begin{table}[ht]
\centering
\begin{tabular}{ccc}
     \toprule
     Model & Average score & Usage of keywords\\
\midrule
GPT-3.5 Turbo & \textbf{4.0 (Likely)} & 46.55 \%  \\
GPT-4 Turbo & 3.8 (Likely) & \textbf{54.75 \%}   \\ 
\bottomrule
\end{tabular}
\caption{Summary of the Likert scale and free text re-formulation results.}
\label{tab:gpt-likert}
\end{table}

\subsection{Discussion and Limitations}
The results of our user study highlight several key insights into the evolving capabilities and user interactions with state-of-the-art large language models, particularly GPT-3.5 and GPT-4.

\begin{itemize}
 \item \textbf{Improved Accuracy:} The considerable improvement in intent recognition accuracy from GPT-3.5 to GPT-4 indicates a substantial advancement in the model's ability to understand and categorize user intent. This is particularly notable in categories such as `factual queries', `explanatory inquiries', and `content creation', where GPT-4 demonstrates a significant lead over its predecessor.

\item \textbf{Consistent Challenges in Certain Categories:} Despite these advancements, both models exhibit consistent difficulties in certain intent categories that occur less frequently, such as `curricular planning' or  `learning support'. For these kinds of categories, GPT-3.5 often outperforms GPT-4. This suggests inherent challenges in these areas that future developments of LLMs might focus on.

\item \textbf{Unbalanced Data Distribution:} Our study indicates that ChatGPT users are typically driven by informational intents rather than creative, exploratory, or planning intents such as `artistic exploration' or `curricular planning', regardless of the nature of the preceding chat history. 

\item \textbf{Varied User Preferences and Subjectivity:} The user preference for answers to reformulated prompts of GPT-4 over GPT-3.5, and the overall preference for model answers to original prompts, also highlights the subjectivity in user satisfaction. It suggests that while accuracy and intent recognition are important, they are not the sole determinants of user satisfaction, highlighting thus the necessity for further research in this area.

\item \textbf{The Curse of Persuasive LLM Answers:} Users interacting with GPT models on OpenAI's platform are cautioned: ``ChatGPT can make mistakes. Consider checking important information''. Accordingly, for factual queries, our template encouraged models to include references to aid user fact-checking. Our findings reveal that participants favored answers to their original prompts even if they did not contain references when using the more advanced GPT-4 model, while for the earlier GPT-3.5 model, the answers to reformulated prompts were preferred. This suggests that advanced models like GPT-4 can satisfactorily and convincingly answer user prompts, even without references.

\end{itemize}

\textbf{Implications for LLM Development:} Our results demonstrate the rapid pace of improvement in LLMs, suggesting a trajectory towards even more specialized and accurate models. The discrepancies in performance across different intent categories could guide future developments.
In addition, the mixed user preferences regarding answers to their original prompts versus those resulting from intent-based reformulations highlight the importance of educating users on effective prompt formulation and a more objective analysis of the provided answers. In this context, our study indicates a user willingness to learn and adapt to this technology.\\

\textbf{Limitations and Future Research Directions:}  The observation that users often favored original answers raises questions about the ability of users to discern the accuracy of LLM outputs. This is particularly crucial given the risk of misinformation, as noted in cases where GPT models could provide incorrect information. Furthermore, the uneven data distribution among different intent categories and the noted biases (e.g., in cases of explanatory approaches in GPT-4) suggest a need for more balanced and controlled datasets in future studies. In addition, the consistent difficulty in certain intent categories like `learning support' points to the need for a deeper examination and possible refinement of these categories, potentially splitting them into more nuanced sub-categories.

\section{Conclusion}
\label{sec:conclusion}
In this work, we analyzed the capabilities of GPT-3.5 Turbo and GPT-4 Turbo to recognize user intents and the effect of intent-based prompt reformulation on user satisfaction. Our study revealed that while ChatGPT models are considerably improving at recognizing user intents, they still struggle to recognize unusual intents. Interestingly, for some unusual intent categories, GPT-3.5 outperforms GPT-4 regarding recognition rates. Furthermore, the study showed that whenever the intent is correctly recognized for both models, the users still prefer the answers to their original prompts to those from the intent-based reformulations. This finding is surprising and contradicts the current literature on prompt reformulation. Despite their potential benefits for unusual intent categories, strategies akin to chain-of-thought-like prompt reformulation seem to become less impactful with model improvements, as indicated by the results of this study.

\bibliographystyle{unsrtnat}
\bibliography{references.bib}

\newpage
\section{Appendix}
\subsection{Detailed Taxonomy}

\begin{longtable}{>{\centering\arraybackslash}m{1cm}@{\hspace{5pt}}m{14.5cm}}
\toprule
\multicolumn{1}{>{\centering\arraybackslash}m{2cm}}{\textbf{Intent Type}} &
\multicolumn{1}{m{13.5cm}}{\textbf{Fine-granular Intent}} \\
\midrule
\endfirsthead

\begin{sideways} \thead{ Informational Intent \citep{broder2002taxonomy, jansen2008determining, kellar2007field} \\ \citep{rose2004understanding, yi2008deciphering}}\end{sideways} & 
    \begin{itemize}
        \item \textbf{Factual Queries:} Requests for specific facts or data.
        \item \textbf{Explanatory Inquiries:} Explanations or clarifications about concepts, events, phenomena, procedures.
        \item \textbf{Tutorial Requests:} Step-by-step instructions or guidance.
        \item \textbf{Definition Queries:} Looking for definitions of terms or concepts.
        \item \textbf{Biographical Queries:} Information about a person's life.
        \item \textbf{Historical Queries:} Seeking historical events or timelines.
        \item \textbf{Geographical Queries:} Information about locations or places.
        \item \textbf{Statistical Queries:} Looking for statistical data.
        \item \textbf{Advice and Tips:} Seeking recommendations or best practices.
        \item \textbf{Legal Information:} Seeking legal definitions or regulations.
        \item \textbf{Cultural and Social Information:} Understanding customs or social norms.
        \item \textbf{News and Current Events Intent:} Users seek information on recent events or news stories, such as \textit{Breaking News, Political News, Economic Updates, Technology News, Entertainment News, Sports Updates, Medical News}.
        \item \textbf{Health and Wellness Intent:} Users are focusing on general health, fitness, and wellness information, such as \textit{Medical Conditions, Nutrition and Diet, Fitness Programs, Mental Health Resources, Healthcare Providers}.
        \item \textbf{Weather and Environmental Intent:} Users are interested in weather or environmental conditions, such as \textit{Current Weather, Forecasts, Environmental Concerns, Natural Disasters}.
        \item \textbf{Calculation and Conversion Intent:} Users need to perform calculations or unit conversions, such as \textit{Mathematical Calculations, Unit Conversions, Currency Conversions, Time Zone Conversions, Date Calculations}.
        \item \textbf{Scientific and Technical Intent:} Users are seeking specialized scientific or technical information, theories, or explanations, such as \textit{Research Data, Technical Specifications, Engineering Concepts, Scientific Formulas}
        \item \textbf{Navigational Intent:} Users aim to reach a particular website or page, such as \textit{Homepage Access, Subpage Access}
    \end{itemize}\\
\hline
\begin{sideways} \thead{ Problem-Solving Intent \\ \citep{rose2004understanding}} \end{sideways} & 
    \begin{itemize}
        \item \textbf{Decision Support:} Assistance in decision-making through insights, comparisons, evaluations
        \item \textbf{Planning and Organization:} Aid in planning events, organizing tasks, or managing projects
        \item \textbf{Technical Support Intent:} Users are seeking assistance with technical issues, such as \textit{Troubleshooting Guides, Error Codes, Device Manuals, Software Help, Customer Support Contact}.
        \item \textbf{Legal and Regulatory Intent:} Users are looking for legal information or regulatory guidelines, such as \textit{Legal Definitions, Regulatory Compliance, Law Services, Government Policies, Case Law and Precedents}
        \item \textbf{Health and Wellness Intent:} Users are focusing on their own health, fitness, and wellness information, such as \textit{Symptoms Queries, Treatment Queries}.
        \item \textbf{Job and Career Intent:} Users are focused on employment and career opportunities, such as \textit{Job Searches, Resume and CV Writing, Interview Preparation, Career Advice, Professional Networking}.
        
        \item \textbf{Financial Intent:} Users are dealing with monetary matters, such as \textit{Banking Services, Investment Information, Tax Information, Loan and Mortgage Queries, Insurance Services, Cryptocurrency}.

    \end{itemize}\\
\hline
\begin{sideways} \thead{ Creative Intent \citep{kellar2007field} \\ \citep{rose2004understanding, yi2008deciphering}} \end{sideways} & 
    \begin{itemize}
        \item \textbf{Artistic Exploration:} Exploration of artistic styles, techniques, historical art contexts
        \item \textbf{Entertainment and Leisure Intent:} Users are seeking entertainment-related content, such as \textit{Movies and TV Shows, Music and Audio, Books and Literature, Games and Apps, Event Information, Horoscopes and Astrology}.
        \item \textbf{Leisure and Hobby Intent:} Users are engaging in hobbies or leisure activities, such as \textit{Recipes and Cooking, DIY Projects, Gardening Tips, Photography Guides, Collectibles and Antiques}
        \item \textbf{Travel and Tourism Intent:} Users are planning trips or seeking travel information, such as \textit{Destination Research, Flight and Hotel Bookings, Travel Tips and Advice, Visa and Passport Information, Local Customs and Etiquette}.
       \item \textbf{Multimedia Intent:} Users are looking for various types of media content, such as \textit{Image Searches, Video Searches, Audio Searches, Live Streams, Games and Interactive Content, Animation and Graphics}.
    \end{itemize}\\
\hline
\begin{sideways} \thead{ Educational Intent \\ \citep{rose2004understanding}} \end{sideways} &
    \begin{itemize}
        \item \textbf{Learning Support:} Assistance or guidance with understanding educational material, concepts, academic research, theses or dissertation. 
        \item \textbf{Educational Resources:} Looking for learning materials.
        \item \textbf{Skill Development:} Guidance on developing specific skills or competencies, such as \textit{Language Learning, Exam Planning}
        \item \textbf{Curricular Planning:} Help in designing or choosing educational curricula or courses.
        \item \textbf{Organizational and Educational Institutional Intent:} Users are interacting with organizations or institutions, such as \textit{Institute Information, Policy and Procedure Access, Application Requirements, Institutional Announcements}
    \end{itemize}\\
\hline
\begin{sideways} \thead{ Personal Interaction Intent \citep{rose2004understanding} \\ \citep{yi2008deciphering}} \end{sideways} & 
    \begin{itemize}
        \item \textbf{Conversational Engagement:} Dialogue for entertainment, companionship, interaction, finding Easter eggs, hidden features, random facts or trivia.
        \item \textbf{Self-Help Intent:} Users aim to improve personal skills or well-being, such as \textit{Self-Improvement Guides, Self-Reflection, Personal Growth, Career Development, Financial Planning}.
        \item \textbf{Social and Community Intent:} Users aim to engage with social platforms or communities, such as \textit{Social Networking, Community Forums, Q\&A Participation, Event Participation, Content Sharing}.
        \item \textbf{Personal Intent:} Users aim to improve personal skills or well-being, such as \textit{Relationship Advice, Lifestyle Changes}.
        \item \textbf{Contextual and Personalization Intent:} Users are interacting with personalized or context-aware services, such as \textit{Personal Schedules, Reminder Settings, Customized Content, Voice Assistant Commands}.
        \item \textbf{Local Intent:} Users are seeking information relevant to a specific geographic area, such as \textit{Local Business Searches, Service Providers, Directions and Maps, Local Events, Public Services, Traffic Updates, Weather Information, Real Estate Searches}.
        \item \textbf{Temporal Intent:} Users are seeking time-sensitive information, such as \textit{Event Timings, Deadlines and Cut-offs}.
        
    \end{itemize}\\
\hline
\begin{sideways} \thead{ Technical and Professional Intent \citep{broder2002taxonomy} \\ \citep{jansen2008determining, rose2004understanding, yi2008deciphering}} \end{sideways} & 
    \begin{itemize}        
        \item \textbf{Technical Support Intent:} Users are seeking assistance with technical issues, such as \textit{Troubleshooting Guides, Error Codes, Device Manuals, Software Help, Customer Support Contact, Professional Fields}.
        \item \textbf{Legal and Regulatory Intent:} Users are looking for legal information or regulatory guidelines, such as \textit{Legal Definitions, Regulatory Compliance, Law Services, Government Policies, Case Law and Precedents}.
        \item \textbf{Job and Career Intent:} Users are focused on employment and career opportunities, such as \textit{Job Searches, Resume and CV Writing, Interview Preparation, Career Advice, Professional Networking, Professional Development}.
        \item \textbf{Organizational and Institutional Intent:} Users are interacting with organizations or institutions, such as \textit{Company Information, Policy and Procedure Access, Employee Resources, Institutional Announcements}
        \item \textbf{Navigational Intent:} Users aim to reach a particular website or page, such as \textit{Homepage Access, Subpage Access, URL Queries, Brand or Service Access, Person-Specific Pages}.
        
    \end{itemize}\\
\hline
\begin{sideways} \thead{ Transactional Intent \citep{broder2002taxonomy} \\ \citep{jansen2008determining, kellar2007field} \\ \citep{ashkan2009classifying, yi2008deciphering}} \end{sideways} & 
    \begin{itemize}
        \item \textbf{Model Functionalities:} Users seek assistance with specific functions of the model, such as \textit{Service Utilization, Data Processing, Task Automation}.
        \item \textbf{Online Transactions:} Users intend to complete a transaction, such as \textit{E-commerce Transactions (Product Purchases, Service Subscriptions, Price Inquiries), Financial Transactions (Banking Services, Investment Actions), Digital Content Transactions (Software Downloads, Media Streaming, File Downloads)}.
        \item \textbf{Online Service Interaction:} Users intend to interact with online services, such as \textit{Booking and Reservations (Travel Arrangements, Event Tickets, Hotel Reservations), Account Management (Profile Updates, Password Resets), Form Submissions (Applications, Surveys and Feedback)}
        \item \textbf{Commercial Investigation Intent:} Users are researching products or services with potential future transactions, such as \textit{Product Comparison, Reviews and Ratings, Feature Inquiries, Best-of Lists, Price Comparisons, Warranty and Support Information, Discounts and Promotions, Brand Research}.
    \end{itemize}\\
\hline
\begin{sideways} \thead{ Ethical and \\ Philosophical \\ Intent \\ \citep{yi2008deciphering}} \end{sideways} & 
    \begin{itemize}
        \item \textbf{Moral and Ethical Queries:} Questions about ethical dilemmas, moral principles, or philosophical theories.
        \item \textbf{Societal and Cultural Inquiry:} Exploring societal, cultural, or historical topics.
        \item \textbf{Existential Questions:} Delving into existential themes or abstract philosophical questions.
    \end{itemize}\\
\bottomrule
\label{tab:taxonomy_detailed}
\end{longtable}

\subsection{Study design}\label{sec:design}
\begin{figure}[ht]
\centering
\includegraphics[width =0.6\textwidth]{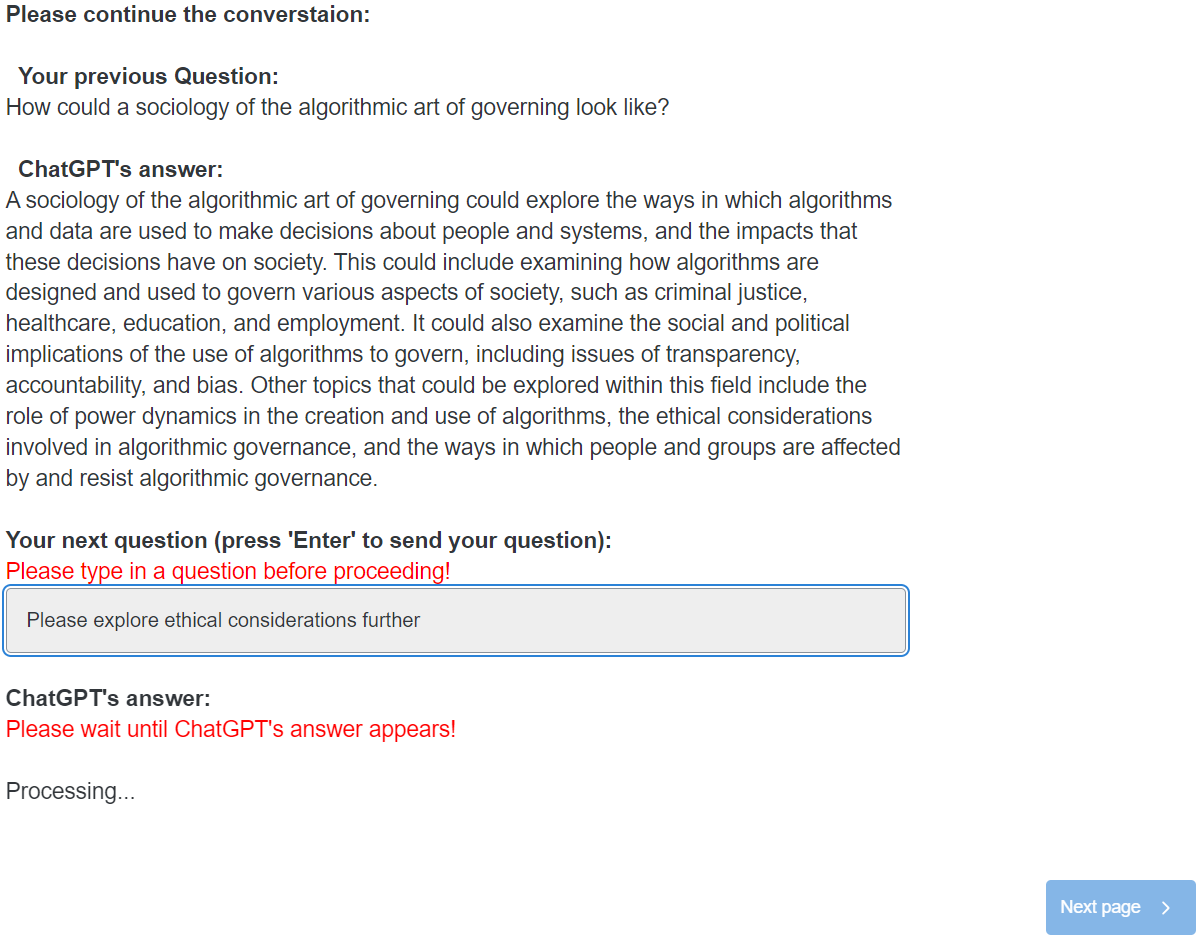}
\caption{Example of the first part of the study}
\label{img:question}
\end{figure}

The users have time to read the randomly selected chat history out of 240 previously selected conversations. Each chat history was previously classified, and 10 samples were saved corresponding to each fine-granular intent categories, to provide diverse answers across the categories. The `Next page' button was only enabled once the GPT answer and the class prediction in the backend was finished (Fig~\ref{img:question}).

\begin{figure}[ht]
\centering
\includegraphics[width =0.6\textwidth]{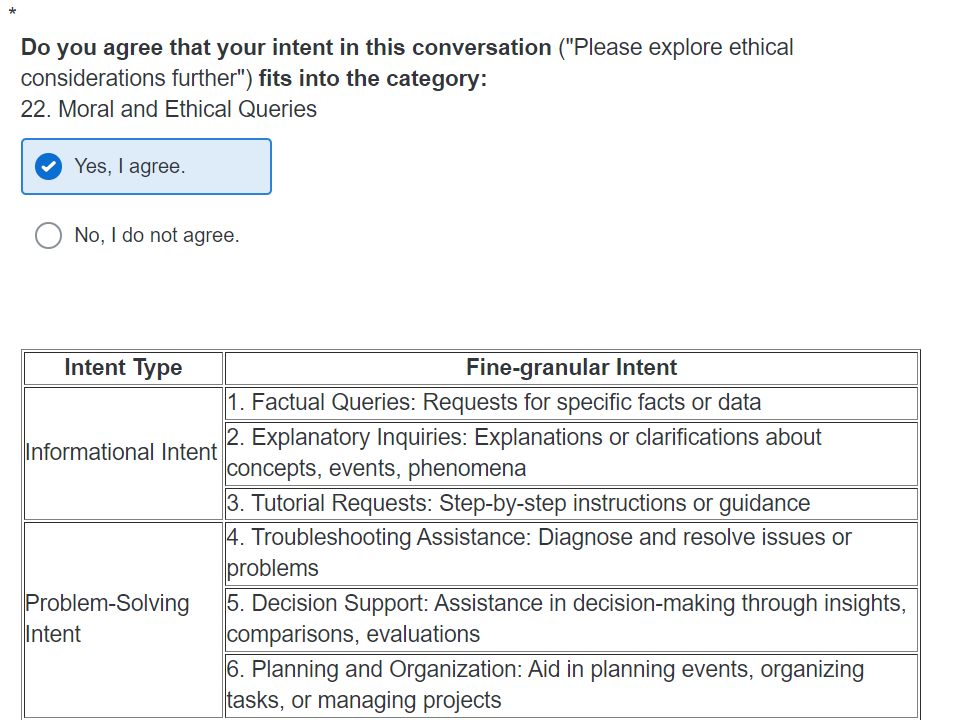}
\caption{Example of intent prediction. For the participants, the complete fine-grained intent categories were provided.}
\label{img:class}
\end{figure}

\begin{figure}[ht]
\centering
\includegraphics[width =0.6\textwidth]{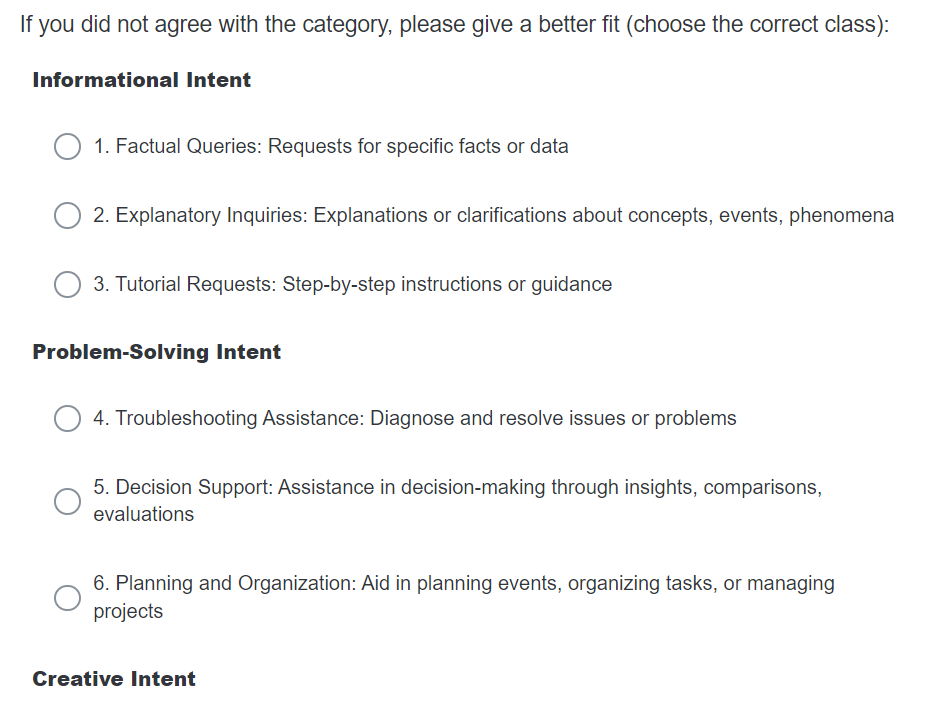}
\caption{Example of intent correction, if the participants did not agree with the predicted intent. The complete list of fine-grained intent was provided.}
\label{img:not_agree}
\end{figure}

In the next page, the participants were asked whether they agree with the predicted intent (Fig~\ref{img:class}). Their previous prompt and a summary table about the fine-grained intent categories were provided to them. If they chose the radio button `Yes, I agree.' they were directed to the next page with a new chat history. In case they did not agree with the predicted intent category, they were directed to a page, where they could select a more suitable intent with the help of radio buttons. Only one intent category could be selected (Fig~\ref{img:not_agree}).

\begin{figure}[ht]
\centering
\includegraphics[width =0.55\textwidth]{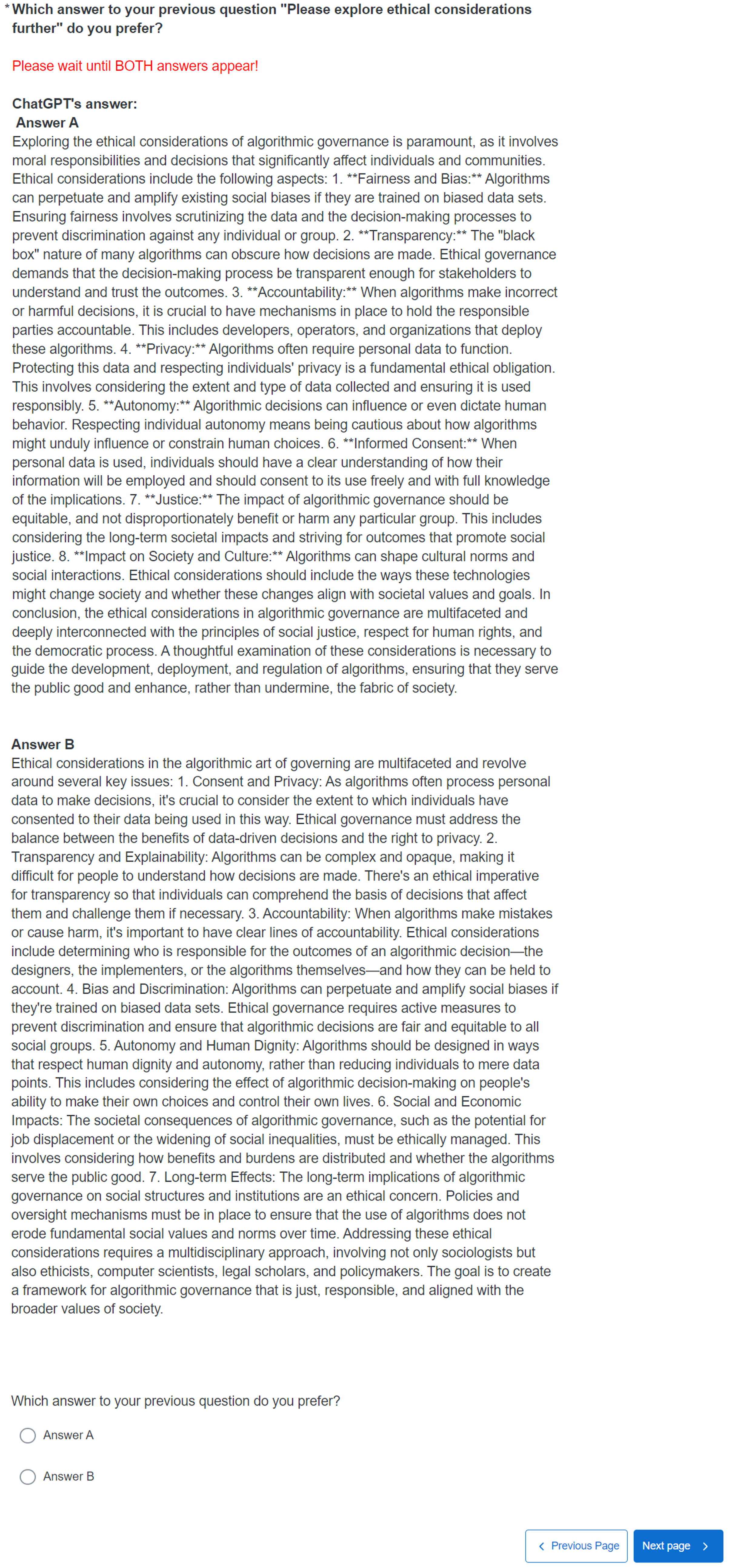}
\caption{Example of the second part of the study}
\label{img:answer}
\end{figure}

After 10 different chat histories were shown and the intent classification was completed, the participants had a chance to chose between the original and re-prompted GPT answers (Fig~\ref{img:answer}). 

\section{User Feedback}
\begin{figure}[ht]
\centering
\includegraphics[width =0.8\textwidth]{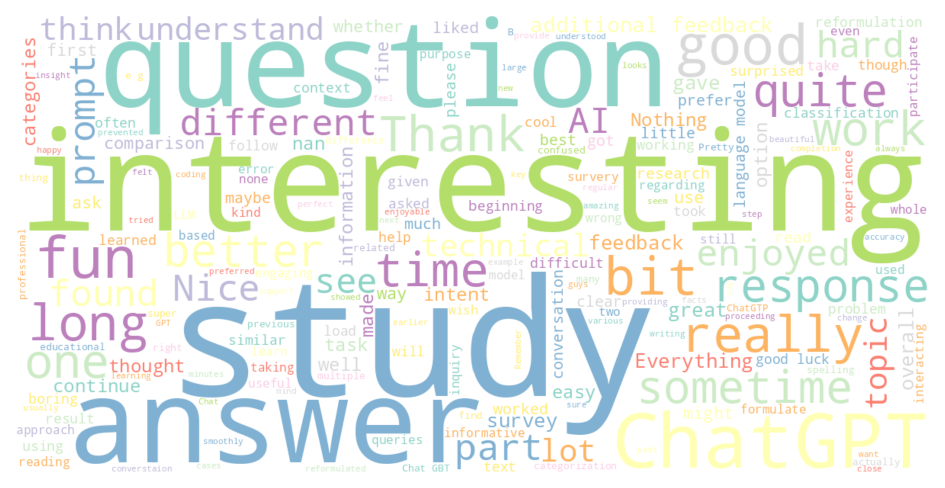}
\caption{Word cloud of the free text received as feedback.}
\label{img:wordcloud}
\end{figure}

We collected feedbacks about the study in a free text format, from where the collected word cloud is visualized in Fig~\ref{img:wordcloud}. In many cases, the participants used this field to say that they enjoyed our study and are curious about the results. Some also mentioned, that they experienced longer waiting times, which could be caused by poor internet connection.

\subsection{Data distribution based on demographics data}
\begin{figure}[ht]
	\centering
	\subfloat[GPT-3.5.]{\includegraphics[height=4.8cm, keepaspectratio]{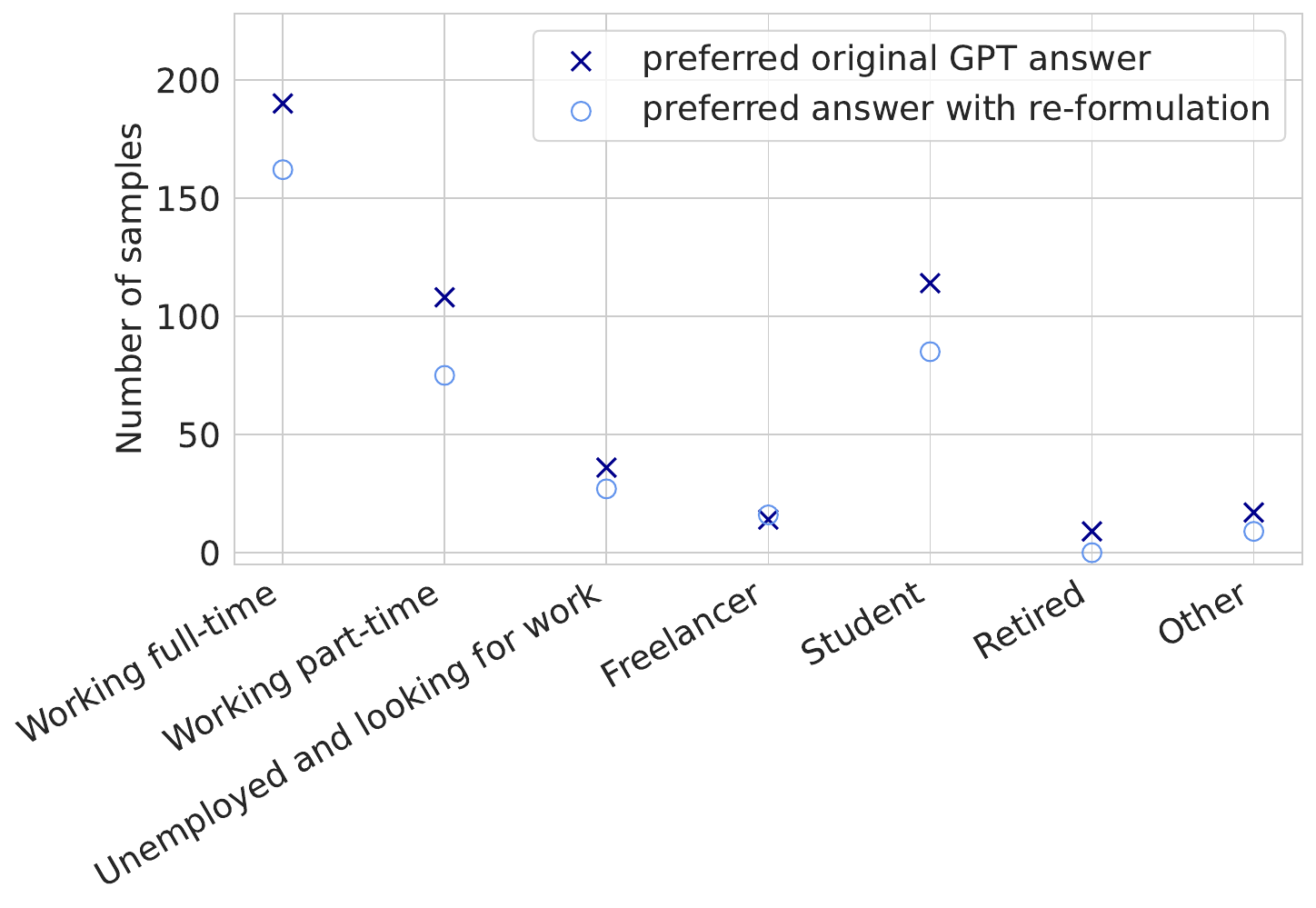}
   \label{fig:GPT-3.5 Turbo_employment}}
	\subfloat[GPT-4.]{\includegraphics[height=4.8cm, keepaspectratio]{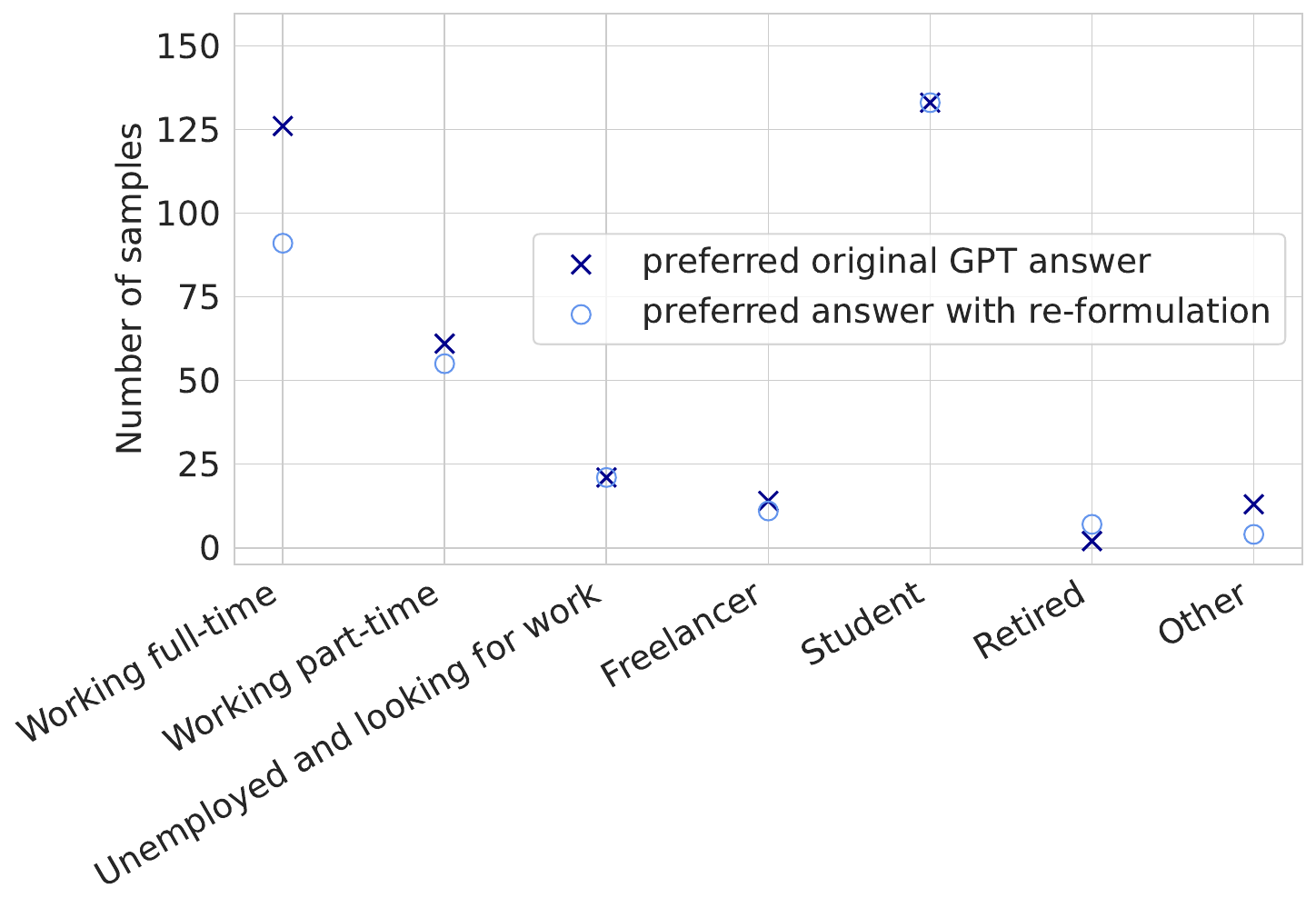}
	\label{fig:gpt4_employment}}
	\hspace{10mm}
	\caption{Data distribution among participants based on their employment status.}
   \label{fig:employment}
\end{figure}

\begin{figure}[ht]
	\centering
	\subfloat[GPT-3.5.]{\includegraphics[height=6.1cm, keepaspectratio]{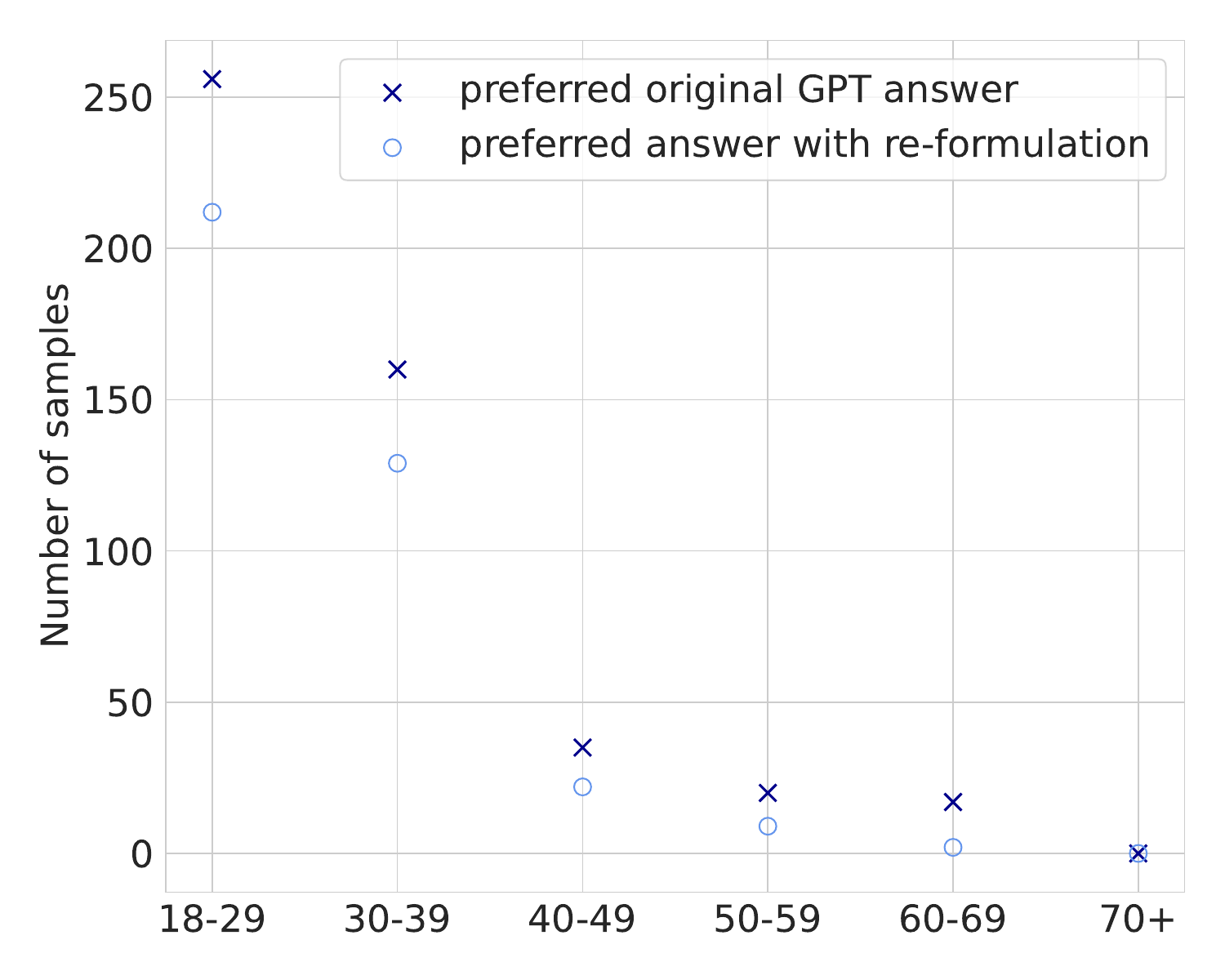}
   \label{fig:GPT-3.5 Turbo_age}}
	\subfloat[GPT-4.]{\includegraphics[height=6.1cm, keepaspectratio]{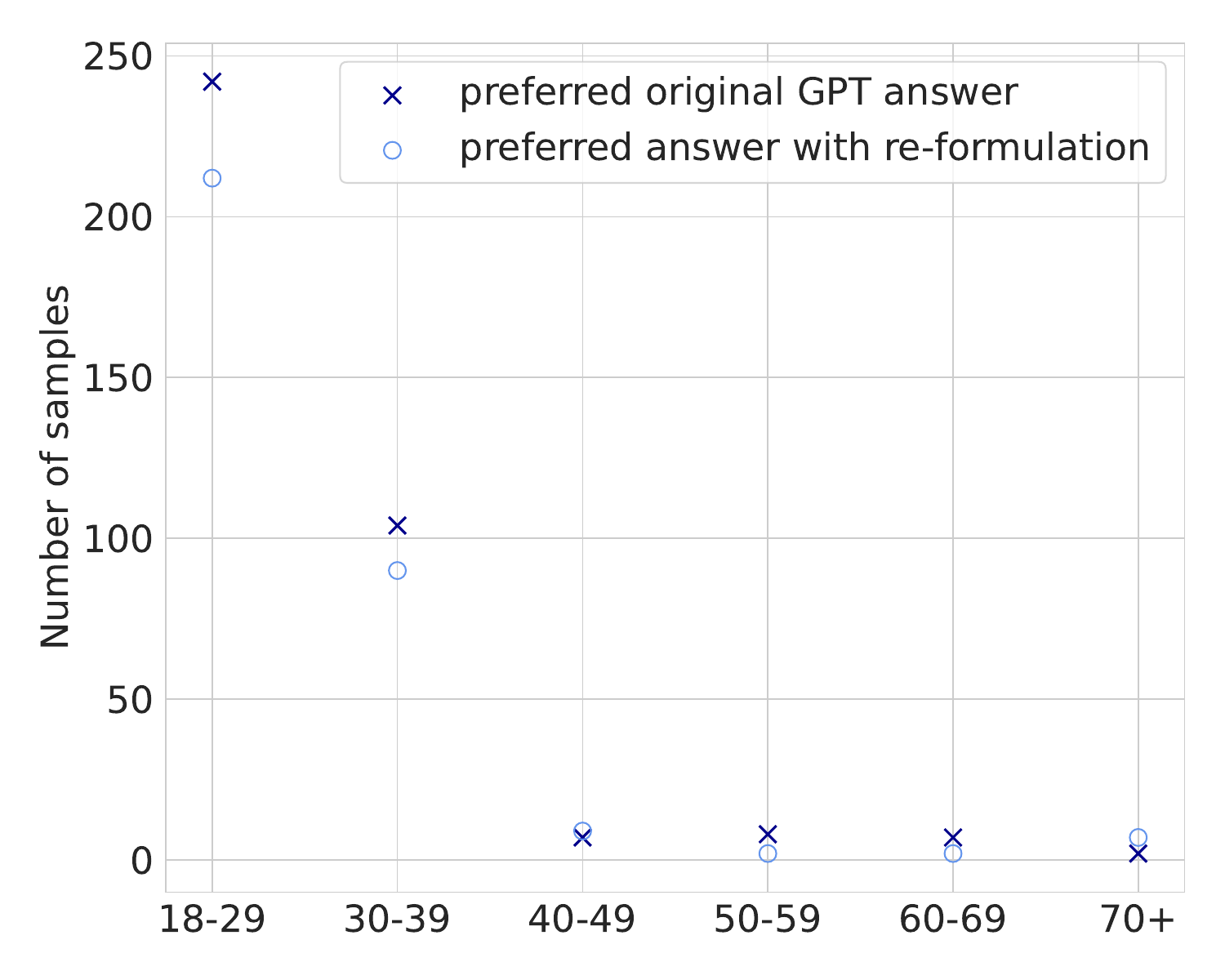}
	\label{fig:gpt4_age}}
	\hspace{10mm}
	\caption{Data distribution among participants based on their age.}
   \label{fig:age}
\end{figure}

We analyzed the collected data samples based on employment status and age, which are demonstrated in Fig~\ref{fig:employment},~\ref{fig:age}. We collected data mostly from participants who were working full-time, working part-time, or students at the time of the study. The majority of the participants below the age of 40. We observed that the ratio between the preferred answers came closer, and the category-wise evaluation shows that in specific categories, users preferred the answers generated with re-prompting.

The same phenomenon can be observed, when we measured the understanding based on the free text reformulation input, where the participants were asked to reformulate the sentence\textit{`Hey, tell me about Albert Einstein. I need info ASAP'}. We filtered out the users, whose answer did not contain the main verbs from the previously provided three templates, namely: \textit{`provide'}, \textit{`assist'}, and \textit{`delve'}. The contrast is even more significant for the GPT-4 model. From these results, we can say that in some cases, users preferred the re-prompted GPT answers, when they understood the main driving force of the study.

\subsection{Evaluation based on user understanding}
We analyzed the collected data based on the understanding of the survey of the participants. Firstly, we filtered the participants, who answered with `Likely' or `Extremely likely' to at least one of the questions asking whether they would use prompt reformulation in the future. 
\begin{figure}[ht]
	\centering
	\subfloat[GPT-3.5.]{\includegraphics[height=6.1cm, keepaspectratio]{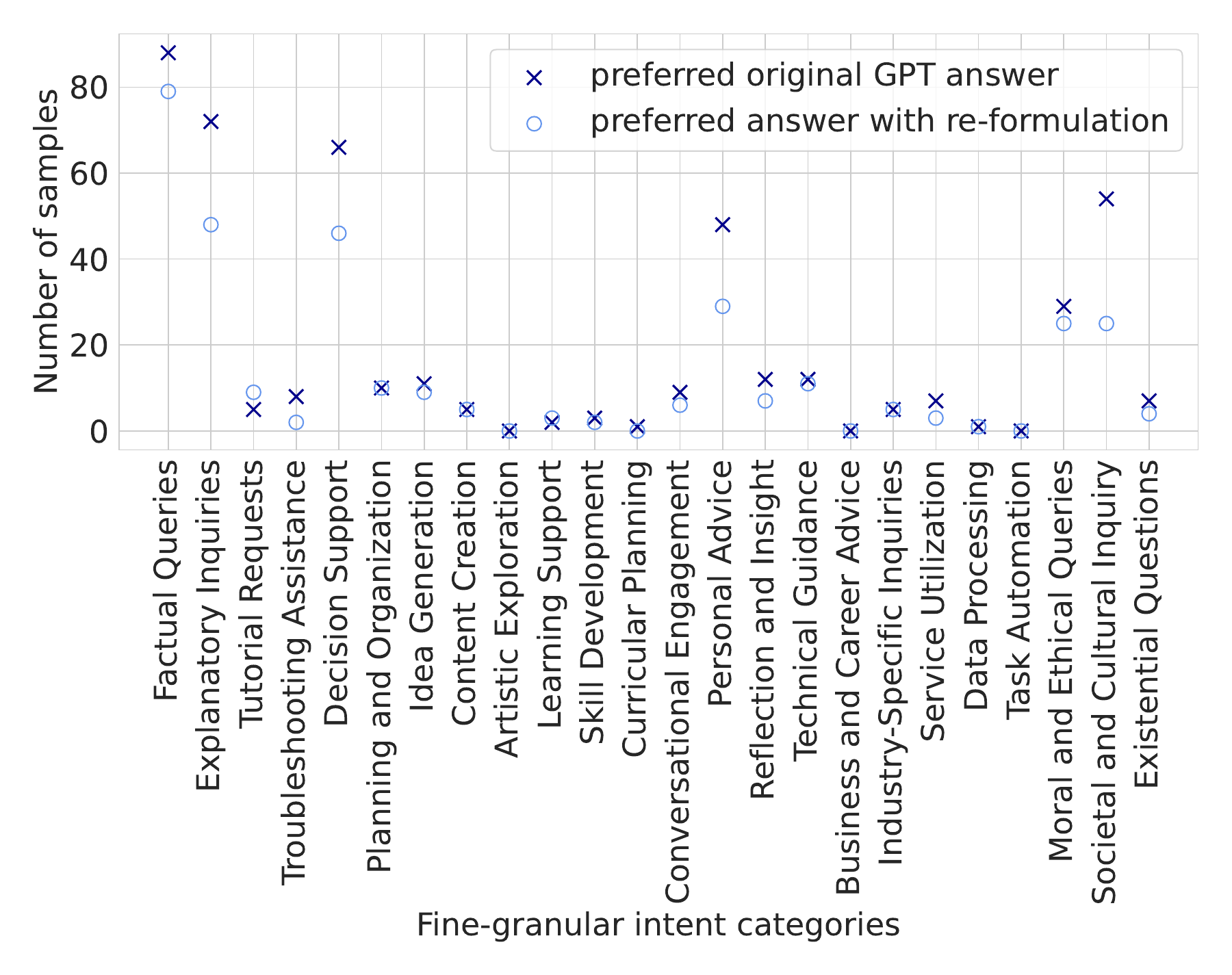}
   \label{fig:GPT-3.5 Turbo_likert}}
	\subfloat[GPT-4.]{\includegraphics[height=6.1cm, keepaspectratio]{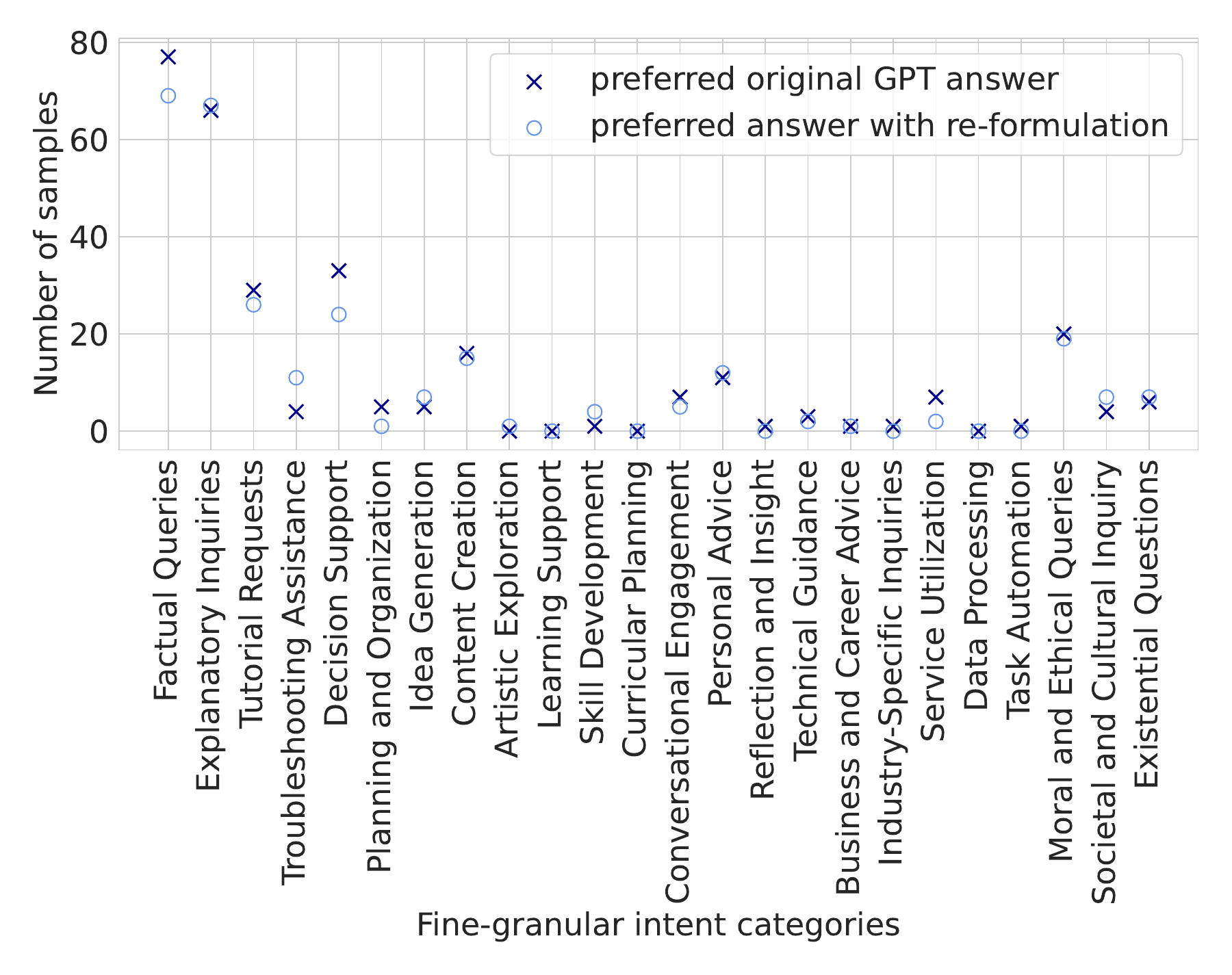}
	\label{fig:gpt4_likert}}
	\hspace{10mm}
	\caption{Results with participants who answered with `Likely' or `Extremely likely' to our questions measuring whether they would use re-formulations.}
   \label{fig:likert}
\end{figure}

\begin{figure}[ht]
	\centering
	\subfloat[GPT-3.5.]{\includegraphics[height=6.1cm, keepaspectratio]{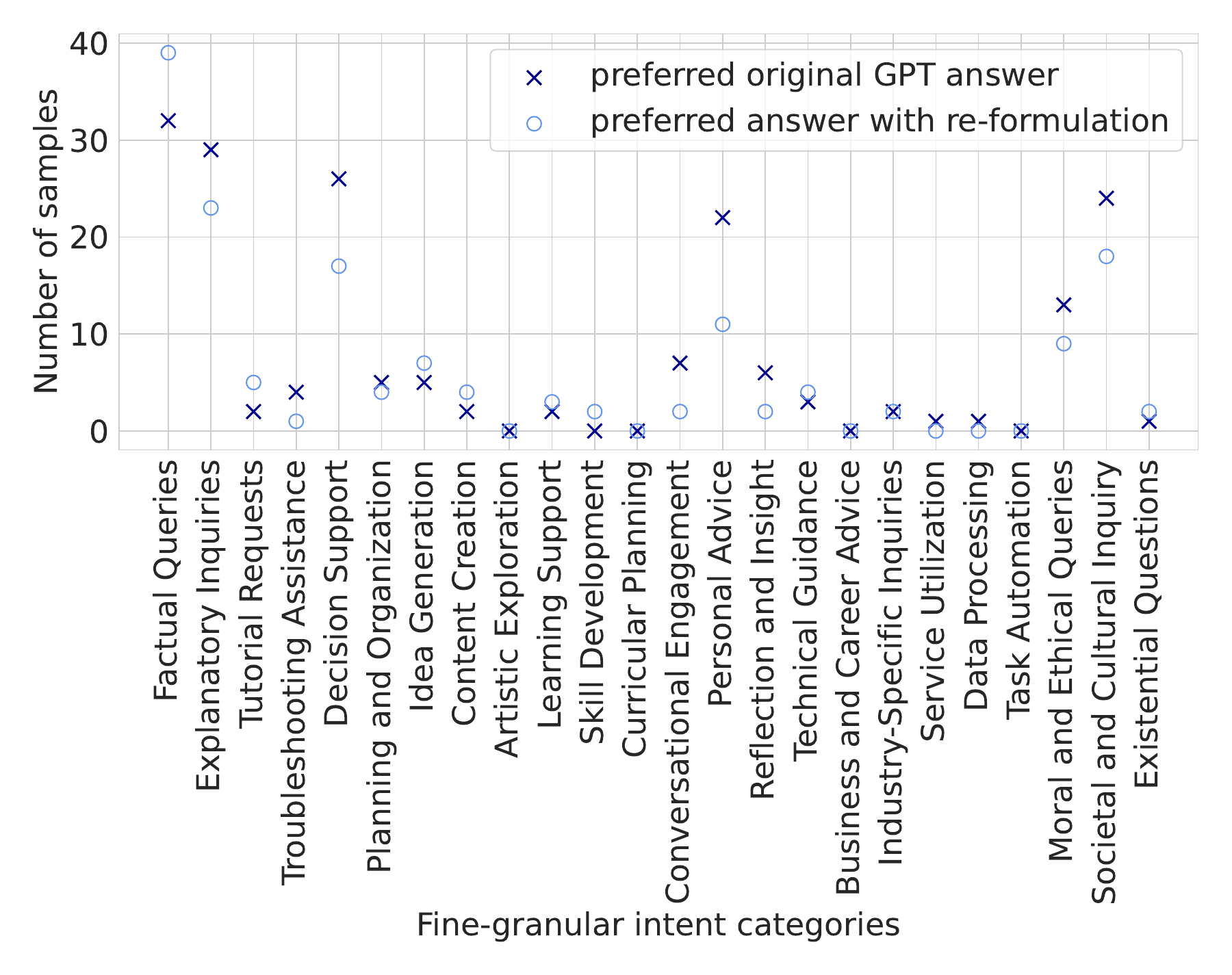}
   \label{fig:GPT-3.5 Turbo_einstein}}
	\subfloat[GPT-4.]{\includegraphics[height=6.1cm, keepaspectratio]{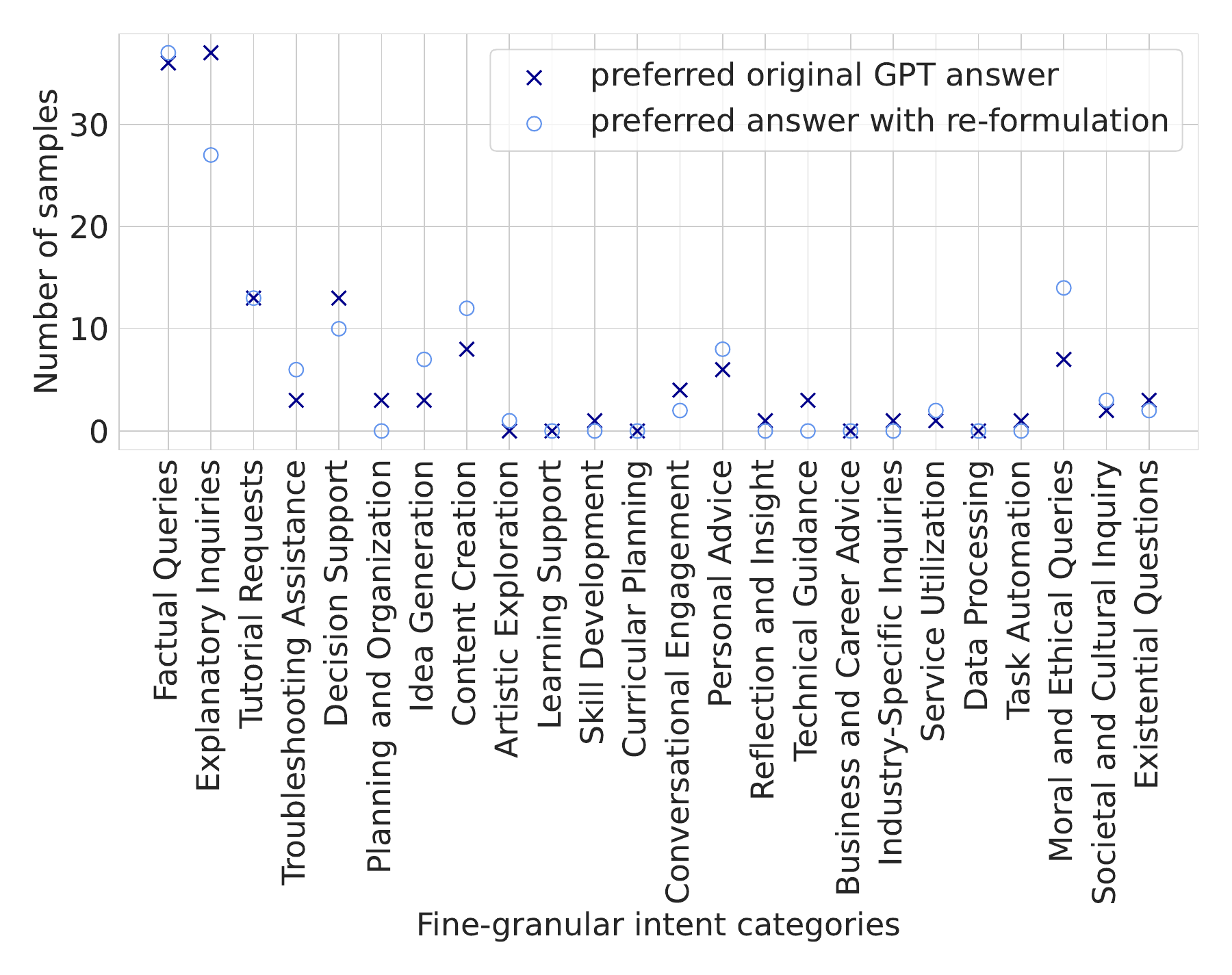}
	\label{fig:gpt4_einstein}}
	\hspace{10mm}
	\caption{Results with participants, who understood templating.}
   \label{fig:einstein}
\end{figure}

\end{document}